\documentclass[preprint2,tighten]{aastex62_reep}

\usepackage{epstopdf}
\usepackage{subfigure}  
\usepackage{amsmath}
\usepackage{hyperref}
\usepackage{graphbox}

\usepackage{longtable}

\usepackage{fancyhdr}   
\fancypagestyle{specialfooter}{%
 \fancyhf{}
 
 \fancyfoot[C]{Distribution Statement A.  Approved for public release: distribution is unlimited.}  }

\shorttitle{Flare Duration in Different Wavelengths}
\shortauthors{Reep \& Airapetian}

\begin{document}

\title{Understanding The Duration of Solar and Stellar Flares at Various Wavelengths}

\author[0000-0003-4739-1152]{Jeffrey W. Reep}
\affiliation{Space Science Division, Naval Research Laboratory, Washington, DC 20375}
\email{jeffrey.reep@nrl.navy.mil}

\author[0000-0003-4452-0588]{Vladimir S. Airapetian}
\affiliation{American University, 4400 Massachusetts Avenue NW, Washington, DC 20016 USA}
\affiliation{NASA/GSFC/SEEC, Greenbelt, MD, 20771, USA}

\begin{abstract}
Recent irradiance measurements from numerous heliophysics and astrophysics missions including SDO, GOES, Kepler, TESS, Chandra, XMM-Newton, and NICER have provided critical input in understanding the physics of the most powerful transient events on the Sun and magnetically active stars, solar and stellar flares.  The light curves of flare events from the Sun and stars show remarkably similar shapes, typically with a sharp rise and protracted decay phase.  The duration of solar and stellar flares has been found to be correlated with the intensity of the event in some wavelengths, such as white light, but not in other wavelengths, such as soft X-rays, but it is not evident why this is the case.  In this study, we use a radiative hydrodynamics code to examine factors affecting the duration of flare emission at various wavelengths.  The duration of a light curve depends on the temperature of the plasma, the height in the atmosphere at which the emission forms, and the relative importance of cooling due to radiation, thermal conduction, and enthalpy flux.  We find that there is a clear distinction between emission that forms low in the atmosphere and responds directly to heating, and emission that forms in the corona, indirectly responding to heating-induced chromospheric evaporation, a facet of the Neupert effect.  We discuss the implications of our results to a wide range of flare energies.
\end{abstract}


\section{Introduction}
\thispagestyle{specialfooter}   

The Sun and all other cool dwarf stars produce powerful eruptive processes via magnetic reconnection in the form of flares and coronal mass ejections (CMEs).  Flares cause the dissipation of magnetic energy, generating emission across the electromagnetic spectrum, from radio to $\gamma$-rays.  The relationship between various emission bands can be used to understand the mechanism of magnetic energy conversion into thermal and kinetic energy, and thus provide insights into the physics of flares.  Studies of the energy partition in solar flares provide an estimate of the total solar irradiance in the ultraviolet-optical bands and X-ray bands. 


Observations of flares with Kepler \citep{borucki2010} and the Transiting Exoplanet Survey Satellite (TESS; \citealt{ricker2014}) on other solar-like stars opened a new avenue in the study of flare physics by discovering frequent superflares, events with energy above $10^{33}$ erg and often over $10^{36}$ erg \citep{maehara2012}, whereas the largest solar events are around a few $\times 10^{33}$ erg  \citep{emslie2012}.  Younger solar analogues in the first 100 Myr of their life produce superflares at a rate as high as 1 event/day (\textit{e.g.} \citealt{colombo2022}).  Eruptions from the Sun and other active stars can generate high fluxes of ionizing radiation in the X-ray and extreme ultraviolet bands that can affect the magnetosphere, upper atmosphere, and climate of our planet as well as impact atmospheric evolution of exoplanetary atmospheres via photoionization driven heating that ignites atmospheric escape and affects atmospheric chemistry \citep{airapetian2017,airapetian2020,siskind2022}.  These effects can influence the planetary climate and produce detrimental conditions for habitability.  The D-region of the ionosphere, in particular, responds to even small scale changes in the X-rays \citep{hayes2017} and in proportion to their brightness \citep{hayes2021}.  Thus, studies of stellar variability hold the key to understanding the potential extremes of solar/stellar flare effects and their impacts on space weather in the solar system and other exoplanetary systems.  The peak intensity of the flare and the duration for which ionizing radiation remains elevated both are important to understand the impact on planetary atmospheres.  However, the mechanisms that drive the energy release and transport in solar and stellar flares are not fully understood.  

In this paper, we focus on one aspect of flare physics, the flare duration.  Solar and stellar flares occur over a range of durations spanning two to three orders of magnitude, from minutes to days.  The measured duration depends on the wavelength in which the flare is observed, and the duration varies for the same event at different wavelengths.  In some wavelengths, there appears to be a correlation between the flare intensity and the duration (\textit{e.g.} white light, \citealt{maehara2015}), while the same relation is not apparent in other wavelengths (\textit{e.g.} X-rays).  Why does flare intensity correlate with duration in some wavelengths, but not others?  

\subsection{Observations}

Solar flares are often categorized by their properties as seen by the Geostationary Orbital Environmental Satellites (GOES), a series of satellites operated by NOAA that has provided continuous coverage of the Sun's X-ray activity since the 1970s.  Specifically, the X-ray Sensors (XRS) onboard GOES are two instruments measuring the solar irradiance in soft X-ray (SXR) bands between 0.5 -- 4 \AA\ (3 -- 24 keV) in XRS-A and between 1 -- 8 \AA\ (1.5 -- 12 keV) in XRS-B.  The peak flux in 1-minute XRS-B data is used to classify solar flares, where X-class flares have flux $\ge 10^{-4}$ W m$^{-2}$, and M-, C-, B-, and A-flares have flux decreasing in order of magnitude, respectively.  \citet{reep2019} showed that the duration of solar flares, as measured by the full-width-at-half-maximum (FWHM) in either XRS channel, is not correlated with the flare's class, and is consistent with a log-normal distribution.  The temperature and emission measure as inferred from XRS also do not appear to be correlated with the duration.  However, \citet{toriumi2017} showed that there is in fact a relation between the magnetic energy in a flare and the duration in XRS ($\tau_{\text{FWHM}} \propto E_{\text{mag}}^{0.45 \pm 0.05}$), but they note that the magnetic energy is not clearly correlated with the peak intensity of the SXR emission, so there is some disconnect between the energy release and resultant intensity.

The duration of flares observed in white light is clearly correlated with the released energy of those events.  For example, \citet{maehara2015} found that there is a correlation between the energy release and duration of superflares observed on solar-like stars with the Kepler satellite in white light emission.  Those authors derived a scaling law relating the duration of a flare to its energy, $\tau_{\text{flare}} \propto E_{\text{flare}}^{1/3}$, while their observations found $\tau_{\text{obs}} \propto E_{\text{obs}}^{0.39 \pm 0.03}$.  \citet{namekata2017} found a similar relation between the duration and intensity of white light emissions in M- and X-class solar flares, although the solar events were systematically of shorter duration than the stellar events, which those authors suggest is evidence that the stellar flares have larger field strengths.  Results from the TESS satellite by \citet{pietras2022} found a similar correlation between duration and intensity, and also examined relations with stellar spectral type and magnetic field strength.  \citet{getman2008} also examined X-ray flares in pre-main sequence stars, finding a similar relationship between the duration and luminosity (Getman 2023, private communication).  

In a set of well-observed large solar flares, \citep{toriumi2017} found that the SXR duration was uncorrelated with active region parameters (sunspot area, total magnetic flux, average magnetic field strength), but that the duration was linearly correlated with properties of the flare ribbons (distance between ribbons, ribbon area, and ribbon magnetic flux).  The authors argue that magnetic reconnection continues for a longer time when the magnetic flux is larger, and thus the energy release is related to the flare duration.  This was explored with modeling of a flare arcade in \citet{reep2017}, who showed that SXR duration depends directly on the duration of magnetic reconnection as new loops are energized in turn, as well as on the loop length of the longest loops in the arcade, which dictates the approximate cooling time.  Additionally, the occurrence of an associated CME does not correlate with SXR duration \citep{harra2016}; that is, eruptive and non-eruptive flares are not distinguishable by duration.  Finally, \citet{hayes2020} found a correlation between the SXR duration and the period of quasi-periodic pulsations (QPPs) in the same energy band, and that the QPP period is also correlated with the ribbon properties.  Furthermore, the QPP period is independent from the peak GOES/XRS flux, and the relation between QPP period and flare duration scales differently in the presence or absence of a CME.  However, QPPs are not observed in all flares, but their detection is more likely in larger class flares \citep{simoes2015b, hayes2020}.  

The relation of QPP period to flare duration raises another interesting consideration.  While the mechanisms that produce them are not definitively known \citep{zimovets2021}, QPPs are clearly related to the energy release.  The analysis of an X-class flare by \citet{hayes2019} shows a clear example where they begin to occur during the impulsive phase of the flare, that the hard X-ray (HXR) bursts correspond to peaks in the SXR derivatives (a facet of the Neupert effect), and that the HXR intensity peaks are modulated significantly more than the SXR derivative peaks.  A recent study by \citet{lorincik2022} additionally showed that the QPPs seen in GOES/XRS derivatives and EOVSA radio emission from sources above the looptop corresponded to strong redshifts of \ion{Si}{4} emission observed with IRIS, which was likely a direct signature of outflows from the reconnection region.  

The Neupert effect \citep{neupert1968} describes a relation between the SXRs and HXRs, namely that the intensity of the SXRs is proportional to the fluence of the HXRs:
\begin{equation}
    I_{\text{SXR}} \propto \int I_{\text{HXR}}\ dt
\end{equation}
\noindent or, equivalently,
\begin{equation}
    I_{\text{HXR}} \propto \frac{d}{dt}\Big(I_{\text{SXR}}\Big)
\end{equation}
\noindent This relation occurs because the non-thermal HXRs are a direct signature of heating from the electron beam, which induces chromospheric evaporation that more slowly heats and fills the corona, causing the brightening of SXR emission.  \citet{dennis1993} found that the Neupert effect is apparent in around 80\% of flares.  This effect is more general than just a relation between SXR and HXR emission, in that emission that forms lower in the atmosphere brightens in direct response to the heating, and emission in the corona primarily brightens in response to the evaporation process.  

Bursts in HXR emission, due to bremsstrahlung from non-thermal electron beams, are strongly correlated spatially and temporally with transition region (TR) line emission.  This was first discovered with the Solar Maximum Mission \citep{cheng1981,woodgate1983,poland1984}, and has also been verified with more recent comparisons between SDO/AIA and RHESSI \citep{cheng2012,simoes2015} or IRIS and RHESSI \citep{warren2016}.  This correlation occurs because the heating effectively pushes the TR downwards in height, so that the TR lines form in regions of higher density, immediately causing a jump in intensity ($\propto n^{2}$).  

The production of white light emission in flares occurs when there is a significant energy deposition deep in the chromosphere.  The emission is cospatial with the HXR emission in on-disk observations \citep{krucker2011}.  Additionally, in limb flares, the HXR and white light emission has been found to form at the same heights in the atmosphere \citep{martinez2012,krucker2015}, suggesting that the non-thermal electrons of energy $\gtrsim 30$ keV cause the white light emission \citep{krucker2015}.  \citet{watanabe2017} found that there is a statistical difference in the energy deposition rates due to non-thermal electrons between flares with and without white light enhancements, specifically that white light flares have larger deposition rates over a smaller range of time (\textit{i.e.} much more impulsive heating).  Similarly, it is well known that the energy deposition from non-thermal electrons is well correlated with X-ray intensity, as well as the rise in temperature and emission measure (\textit{e.g.} \citealt{warmuth2016a}).  Additionally, using the data set of solar flares from \citet{warmuth2016a}, we find that there is a correlation between SXR duration and non-thermal electron beam parameters (the electron flux, non-thermal power, and low-energy cutoff), which we show in Appendix \ref{app:beams}.  

Additionally, while the acceleration of electrons is relatively well understood, the transport of energy via other mechanisms like magnetohydrodynamic (MHD) waves or ion beams must also play an important role.  In the largest events, particularly superflares with energy above $10^{33}$ erg, the expected energy fluxes of an electron beam become unstable to Langmuir and ion-acoustic waves \citep{drummond1970,zharkova2011}, so that ion beams likely become significantly more important to the energy balance of the chromosphere.  In this work, we only examine heating due to an electron beam, but other heating mechanisms need to be further constrained in the future.

\subsection{Loop Cooling Time}

The duration of an observed flare depends directly on the cooling time of the loops composing that flare.  Coronal loops cool primarily through thermal conduction, radiation, and enthalpy flux (draining).  Early numerical simulations of coronal loops suggested that there is a direct relation between the cooling and draining time of a loop: $T \propto n^{2}$ for the radiative cooling phase of a flare loop \citep{serio1991,jakimiec1992}, where $T$ is the temperature and $n$ the number density.  More generally, the scaling can be written as $T \propto n^{\delta}$, \citep{bradshaw2005}, where $\delta = \gamma -1 + \frac{\tau_{v}}{\tau_{R}}$ relates the timescales of draining $\tau_{v}$ to radiative cooling $\tau_{R}$, and $\gamma$ is the ratio of specific heats, usually taken to be $\frac{5}{3}$.  For adiabatic cooling, $\delta = \frac{2}{3}$, and for radiation-dominated cooling, $\delta \rightarrow \infty$ \citep{bradshaw2010}.  Additionally, a non-uniform cross-sectional area fundamentally alters the rate of cooling due to both thermal conduction and draining, and thus the typical $T \propto n^{2}$ relation does not hold in general \citep{cargill2022,reep2022}.  Rather, draining is suppressed from the loop ($\tau_{v} \rightarrow \infty$), and radiation dominates the cooling until the onset of catastrophic collapse.  The total cooling time is considerably longer as a result.  

\citet{cargill1995} derived a simple expression to approximate the total (evaporative) cooling time of a loop:
\begin{align}
    \tau_{cool} &\approx \frac{5}{3} \tau_{c}^{7/12} \tau_{r}^{5/12} \nonumber \\ 
    \tau_{cool} &\approx \frac{5 k_{B} L^{5/6}}{\chi^{7/12} \kappa_{0}^{5/12} (nT)^{1/6}} \nonumber \\
    \tau_{cool} &\propto \frac{L^{5/6}}{(nT)^{1/6}} 
    \label{eqn:taucool}
\end{align}
\noindent where $\tau_{c}$ and $\tau_{r}$ are the conductive and radiative timescales, $k_{B}$ the Boltzmann constant, $\kappa_{0}$ the coefficient of thermal conductivity, and $L$ the loop half length.  The radiative losses are commonly taken to have a piece-wise power law form, so that the emissivity can be written $\epsilon(T) = \chi T^{\alpha}$, with $\alpha = -\frac{1}{2}$ in the temperature range of interest.  Equation \ref{eqn:taucool} was derived using the Spitzer-H\"arm approximation to thermal conduction in the absence of electric fields \citep{spitzer1953,spitzer1962,braginskii1965}, but it can be modified to account for turbulence.  \citet{bian2016} derived a treatment for the suppression of thermal conduction in the presence of turbulence, and its impact on scaling laws for the cooling of coronal loops was examined by \citet{bradshaw2019}.  \citet{bian2016b} derived a turbulent conductive timescale:
\begin{align} 
\tau_{\text{Turb}} &= \Big(\frac{c_{R}}{\gamma -1}\Big) \frac{L^{2}}{\kappa_{0} \lambda_{T} T^{1/2}} \nonumber \\
\tau_{\text{Turb}} &\propto \frac{L^{2}}{\lambda_{T} T^{1/2}}
\end{align}
\noindent Importantly, this has the same dependence on loop length as the Spitzer-H\"arm case, has a weaker dependence on temperature, and is independent of density.  This can then be substituted to find the approximate total cooling time of a turbulent loop
\begin{align}
    \tau_{cool} &\approx \frac{5}{2} \Big(\frac{2 k_{B}}{\chi}\Big)^{7/12} \Big(\frac{c_{R}}{\kappa_{0} \lambda_{T}}\Big)^{5/12} \frac{T^{2/3} L^{5/6}}{n^{7/12}} \nonumber \\
    \tau_{cool} &\propto \frac{T^{2/3} L^{5/6}}{\lambda_{T}\ n^{7/12}} 
    \label{eqn:tauturb}
\end{align}
\noindent where $c_{R}$ is a constant and $\lambda_{T}$ is the turbulent mean-free path (see \citealt{bradshaw2019}).  The same dependence on loop length is recovered, but the cooling time is more strongly dependent on both the temperature and density, and it increases with temperature.  In the presence of strong turbulence, hotter loops take significantly longer to cool since energy transport out of the corona by thermal conduction is reduced, particularly in less dense loops.  We will show an example of this in Section \ref{sec:loop}.

In this work, we examine how various cooling and heating parameters affect the evolution of Sun-as-a-star time series.  We have run hydrodynamic simulations and forward modeled emission forming across a wide range of temperatures to better understand how durations can be impacted by different cooling processes, loop geometry, and heating parameters.  We find that different spectral lines evolve in two separate manners: those that respond directly to heating, and those that respond indirectly via chromospheric evaporation.  In other words, there is a general Neupert effect for emission that forms in the corona versus emission in the transition region or chromosphere.  

\section{Model}
In order to explore how flare duration depends on wavelength as well as physical parameters, we employ field-aligned hydrodynamic simulations.  We use the HYDrodynamics and RADiation code (HYDRAD\footnote{https://github.com/rice-solar-physics/HYDRAD}), which describes the evolution of plasma constrained to a magnetic flux tube \citep{bradshaw2003}.  The code includes the physics relevant to coronal loops and flares, including radiative losses from the chromosphere through corona, thermal conduction with either the Spitzer-H\"arm approximation or turbulent suppression of conduction, and heating functions due to direct heating or from an electron beam \citep{bradshaw2013, reep2019b}.  The loop geometry is very general, and can include open or closed loops (\textit{e.g.} \citealt{scott2022}), with uniform or expanding cross-sectional area of any specified form.  In this paper, we examine closed loops both with and without area expansion.  

From the simulations, we obtain the temperatures, densities, velocities, and ion fractions as a function of both space and time, so we can synthesize emission that can be used to facilitate comparisons to observations.  In this paper, we synthesize thermal emission using the CHIANTI atomic database \citep{dere1997}, version 10 \citep{delzanna2021}.  The thermal, optically-thin calculation with CHIANTI includes free-free, free-bound, bound-bound, and two-photon continuum emission.  

We assume solar photospheric abundances \citep{asplund2009} in both the radiative losses and forward model, which is consistent with observations of large solar flares \citep{warren2014}.  Recent observations have additionally found that abundances of some elements (with low first ionization potential) vary with time during a flare, from coronal abundances towards photospheric in the impulsive phase, and then back towards coronal during the gradual phase \citep{mondal2021,sylwester2023,nama2023}.  Additionally, abundances can vary spatially \citep{doschek2016} and from flare to flare \citep{sylwester1998}.  In this work, we do not attempt to model variable elemental abundances, which would modify the radiative losses from a given loop and thus directly impact the duration of emission.  This assumption also implies that flares from stars with metallicities different to the Sun would have different durations as loops would cool at different rates.  That is, the cooling rates depend directly on the metallicity of the host star \citep{sutherland1993,washinoue2023}.  For instance, low metallicity stars such as $\epsilon$-Eridani (HD 22049, \citealt{santos2004}) show flares with longer duration than the Sun \citep{coffaro2020}. 
 Additionally, metallicities of such stars can be up to an order of magnitude lower than solar values \citep{santos2004, testa2010}, and so we would expect the radiative cooling rate to be similarly reduced, by a factor of $\approx 4$ from $\log{T} = 4 - 6$ and a factor of $\approx 2$ around $\log{T} = 7$.  We emphasize that the qualitative behavior of the emission is our focus in this paper, rather than exact durations.   

For the synthesis of optically thick lines (H-$\alpha$, Lyman-$\alpha$, \ion{He}{2} 304), we have run a few simulations of HYDRAD using its NLTE chromosphere \citep{reep2019b} and synthesized the emission with the RH1.5D\footnote{https://github.com/ITA-Solar/rh} radiative transfer code \citep{pereira2015}.  In the calculations of X-ray spectra, we additionally synthesize the non-thermal free-free emission directly, using the assumed electron beam parameters and the full Bethe-Heitler cross-section \citep{bethe1934} with the Elwert correction factor \citep{elwert1939} that accurately captures the cross-section across both non-relativistic and relativistic energies \citep{koch1959}.  We additionally synthesize non-thermal free-bound emission for beam electrons recombining onto \ion{Fe}{22} through \ion{Fe}{26} ions, based on the derivation in \citet{brown2008} and corrections in \citet{reep2016b}.  We synthesize all of the emission as irradiance, that is, spatially unresolved or Sun-as-a-star spectra.  See \citet{kerr2022,kerr2023} for an in-depth review of flare models and forward modeling of radiation in recent years. 

\section{Single Loop Evolution}
\label{sec:loop}

\begin{figure*}
    \centering
    \includegraphics[width=0.48\textwidth]{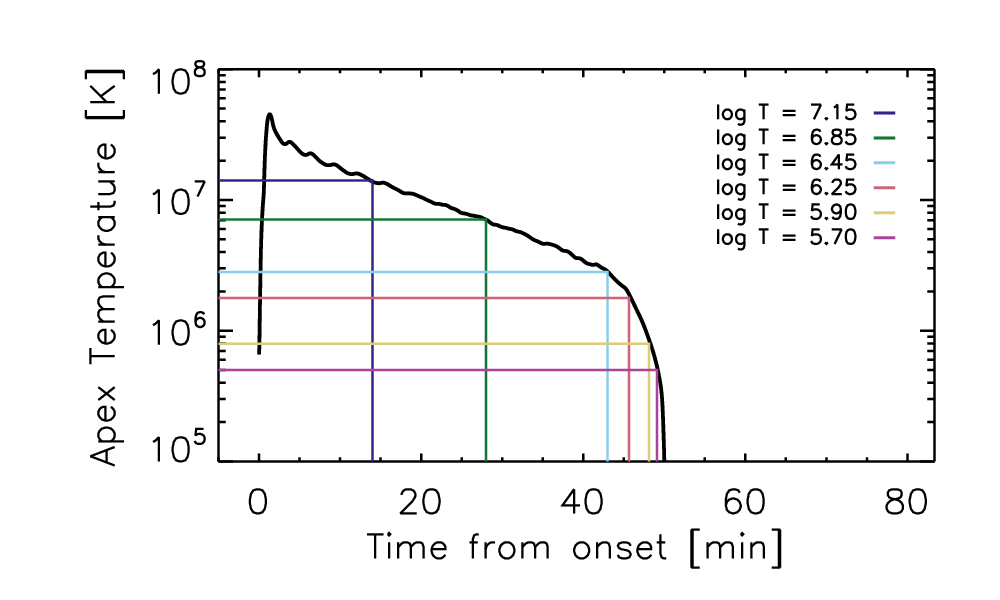}
    \includegraphics[width=0.48\textwidth]{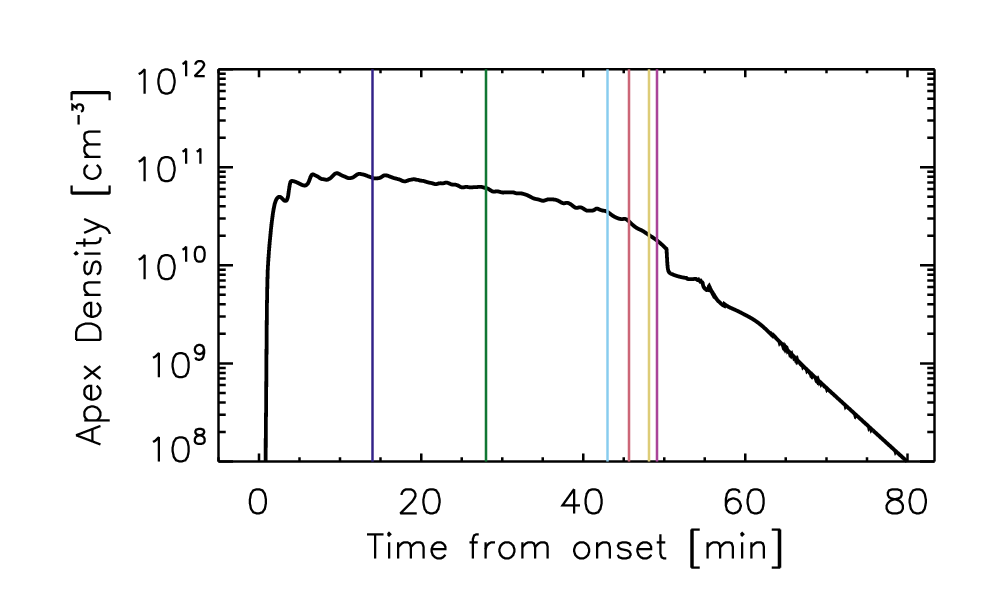}
    \includegraphics[width=0.48\textwidth]{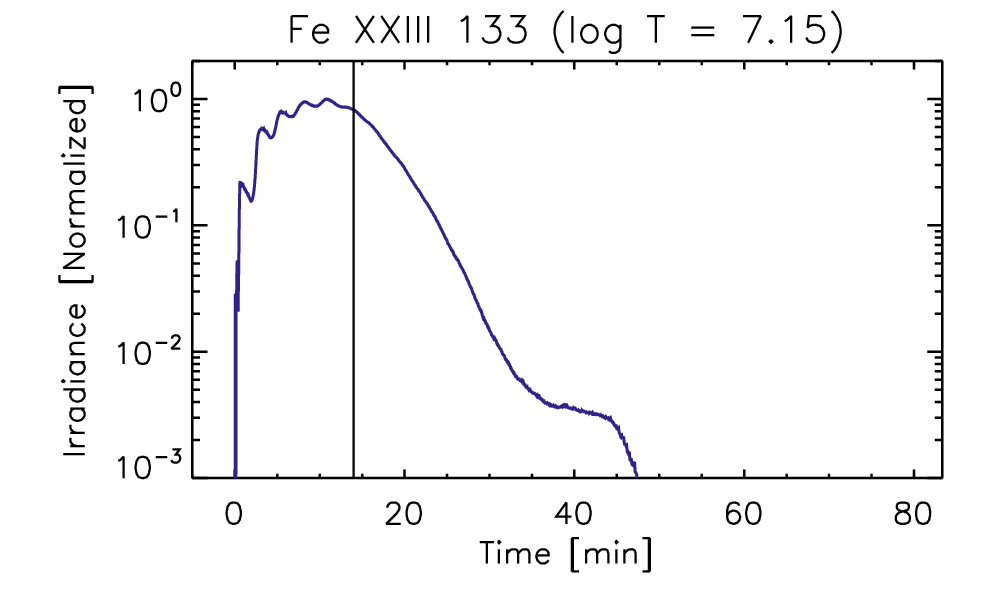}
    \includegraphics[width=0.48\textwidth]{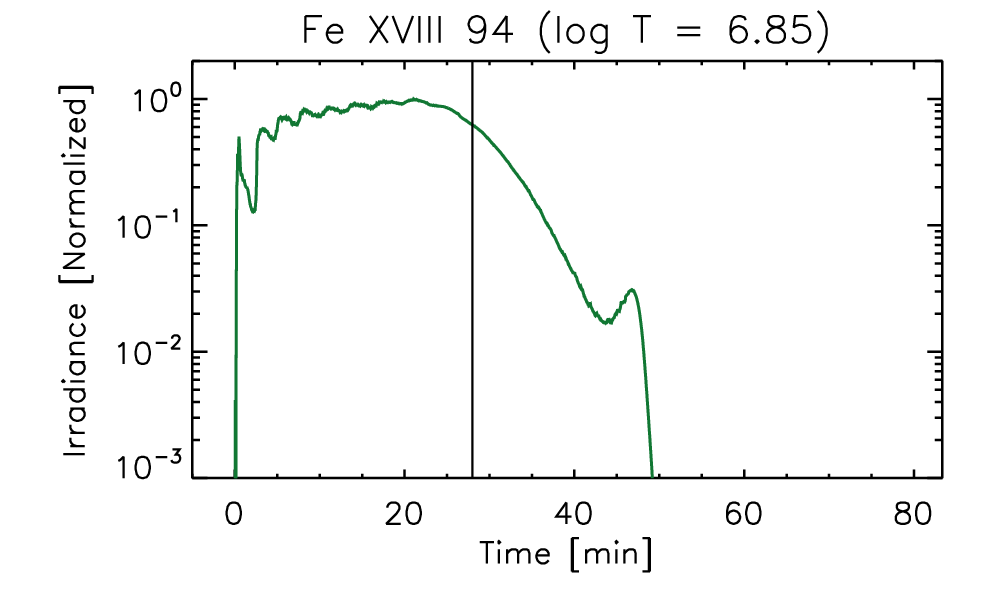}
    \includegraphics[width=0.48\textwidth]{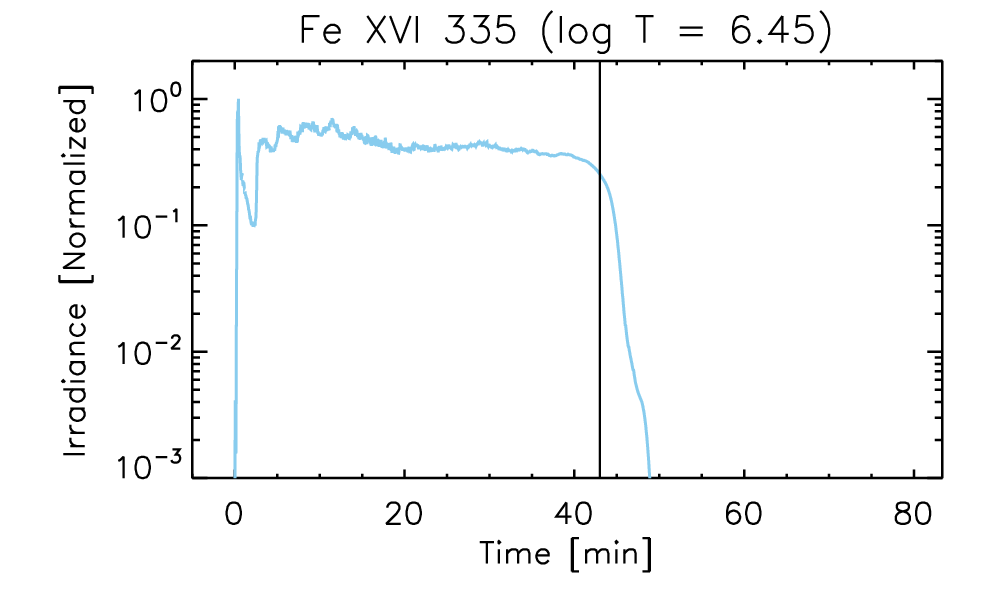}
    \includegraphics[width=0.48\textwidth]{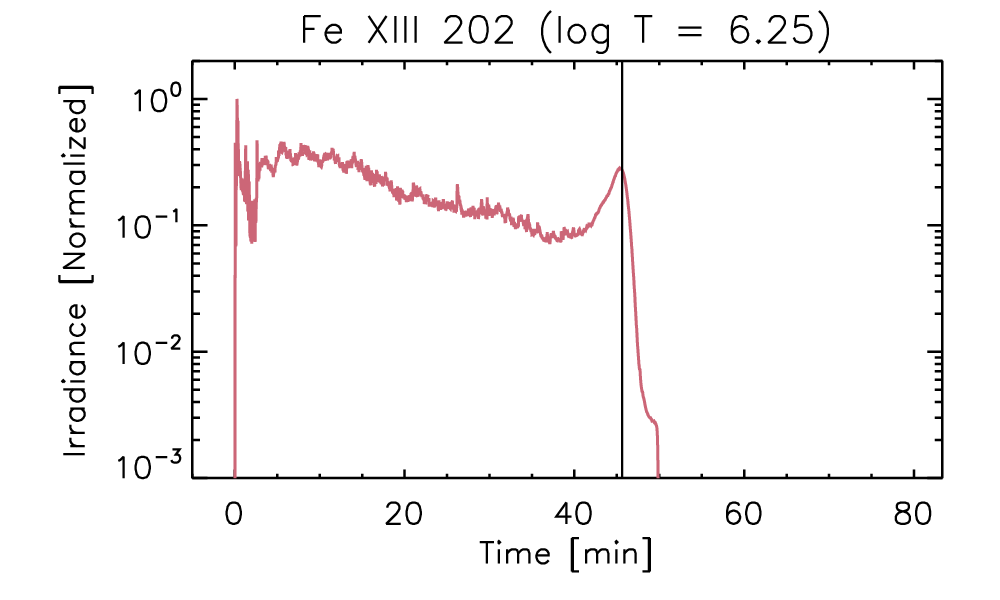}
    \includegraphics[width=0.48\textwidth]{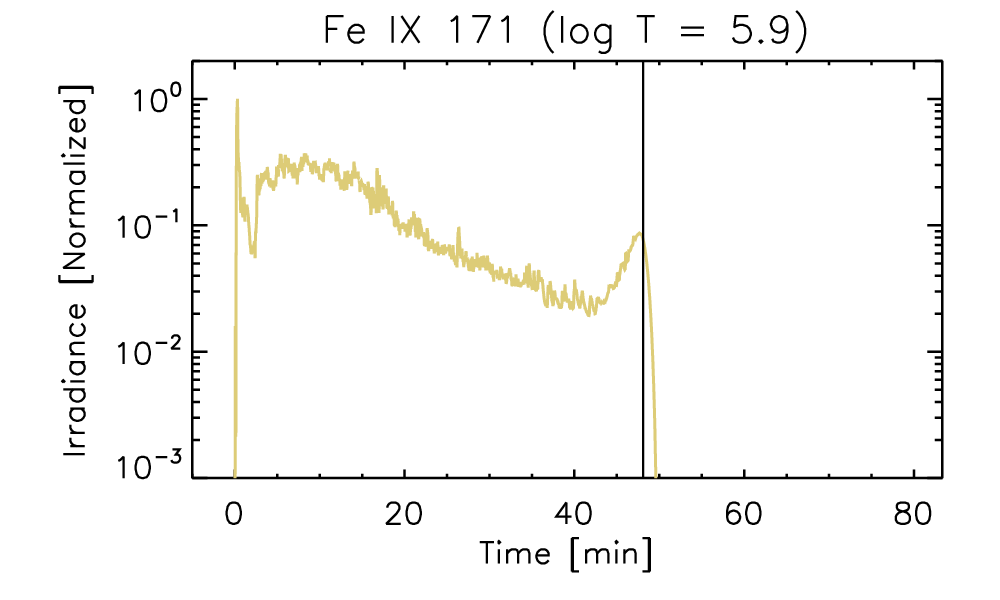}
    \includegraphics[width=0.48\textwidth]{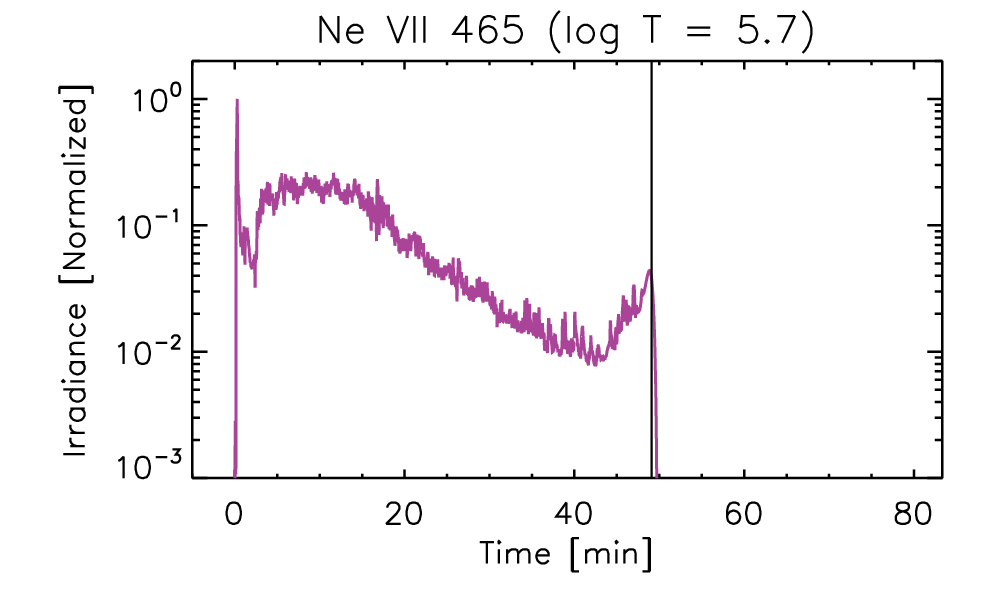}
    \caption{The cooling behavior of a uniform 100 Mm loop heated by a strong electron beam.  The top plots show the time evolution of the electron temperature (left) and electron density (right) at the loop apex.  The bottom six plots show the evolution of synthetic irradiance of 6 spectral lines.  The vertical lines on each plot mark the time at which the coronal temperature reaches the peak formation of each spectral line.  Once the corona cools beyond the formation temperature for each ion, the ion recombines and the irradiance falls off rapidly.  The corona catastrophically collapses after about 50 minutes, and the lines fade entirely after this point.}
    \label{fig:L100}
\end{figure*}

Let us consider first the commonly assumed case of a magnetic loop of uniform cross-sectional area, with the Spitzer-H\"arm approximation to thermal conduction.  Figure \ref{fig:L100} shows an example of the cooling behavior of a 100 Mm loop.  The plasma was heated with an electron beam with a triangular pulse, with $10^{11}$ erg s$^{-1}$ cm$^{-2}$ peak flux, lasting 100 seconds, low-energy cutoff $E_{c} = 25$ keV, and spectral index $\delta = 5$.  The corona is rapidly heated to a temperature above 40 MK, while the density rises more slowly to a peak nearing $10^{11}$ cm$^{-3}$ due to chromospheric evaporation.  The top plots show the evolution of the electron temperature (left) and electron density (right) at the apex of the loop, which takes about 50 minutes in total to cool, and 80 minutes to drain after the onset of heating.  The vertical lines mark the times when the corona hits the peak formation temperature of six spectral lines: \ion{Fe}{23} 133 \AA\ ($\log{T} = 7.15$), \ion{Fe}{18} 94 \AA\ ($\log{T} = 6.85$), \ion{Fe}{16} 335 \AA\ ($\log{T} = 6.45$), \ion{Fe}{13} 202 \AA\ ($\log{T} = 6.25$), \ion{Fe}{9} 171 \AA\ ($\log{T} = 5.90$), and \ion{Ne}{7} 465 \AA\ ($\log{T} = 5.70$).  The bottom six plots show the evolution of the irradiance of these six spectral lines, respectively.  

There is a clear pattern in the time series of these spectral lines in Figure \ref{fig:L100}.  The irradiance of each line rises sharply with the onset of heating, and for the duration of that heating, there is a local peak in the intensity.  This initial spike is most noticeable in the coolest lines, where it is the brightest emission during the entirety of the loop's evolution.  This is because the initial heating pulse pushes the transition region downwards in height, so that the emission of the lines occurs deeper in the atmosphere, where the density is higher.  After that initial spike, there is a gradual increase in intensity that tracks the increase in the density due to chromospheric evaporation.  Then, as the corona cools, the temperature falls below the peak formation temperature of each line in turn (marked by the vertical lines in the plots).  After this occurs, the irradiance of each line falls sharply as the ion population falls due to collisional recombination.  For example, the corona cools to the peak formation temperature of \ion{Fe}{23} at about 14 minutes after heating onset, after which the irradiance of that line falls off rapidly.  In the case of the cooler lines, like \ion{Fe}{9} 171 \AA, the corona is too hot for \ion{Fe}{9} ions to be present before about 40 minutes, so the emission of that line is coming from the transition region prior to then.  As the corona cools below 1 MK, though, the hotter Fe ions recombine to \ion{Fe}{9}, and the line begins to form in the corona itself, causing a spike in the 171 \AA\ line.  Finally, after about 50 minutes, the corona begins to catastrophically collapse, and the emission from all of these lines disappears.

\begin{figure*}
    \centering
    \includegraphics[width=0.48\textwidth]{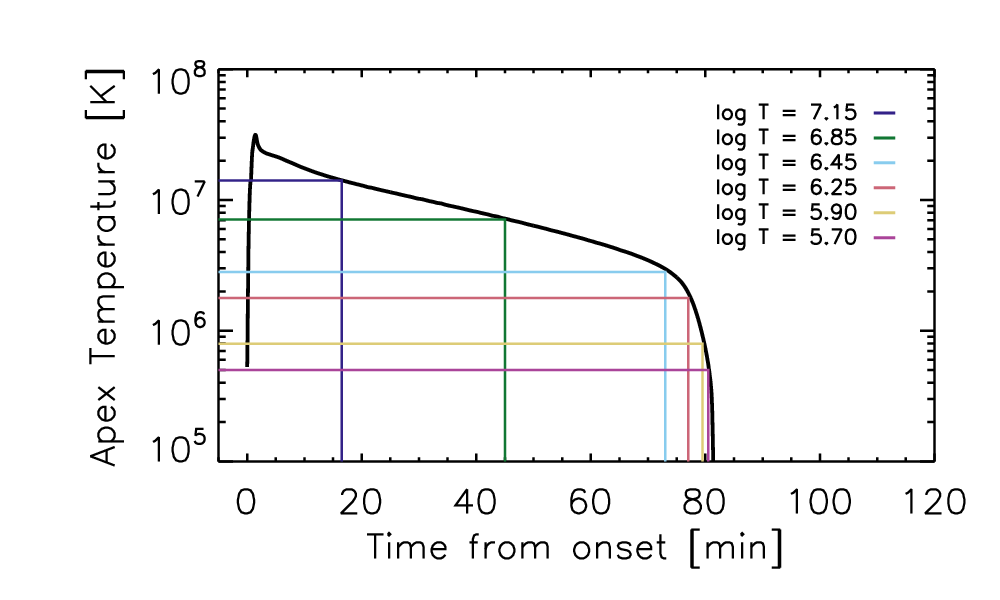}
    \includegraphics[width=0.48\textwidth]{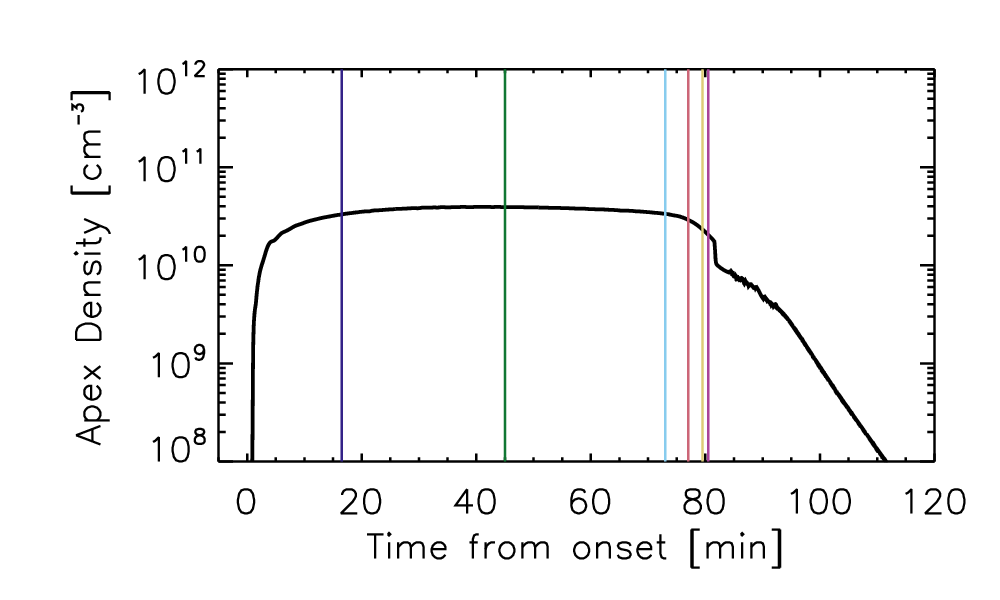}
    \includegraphics[width=0.48\textwidth]{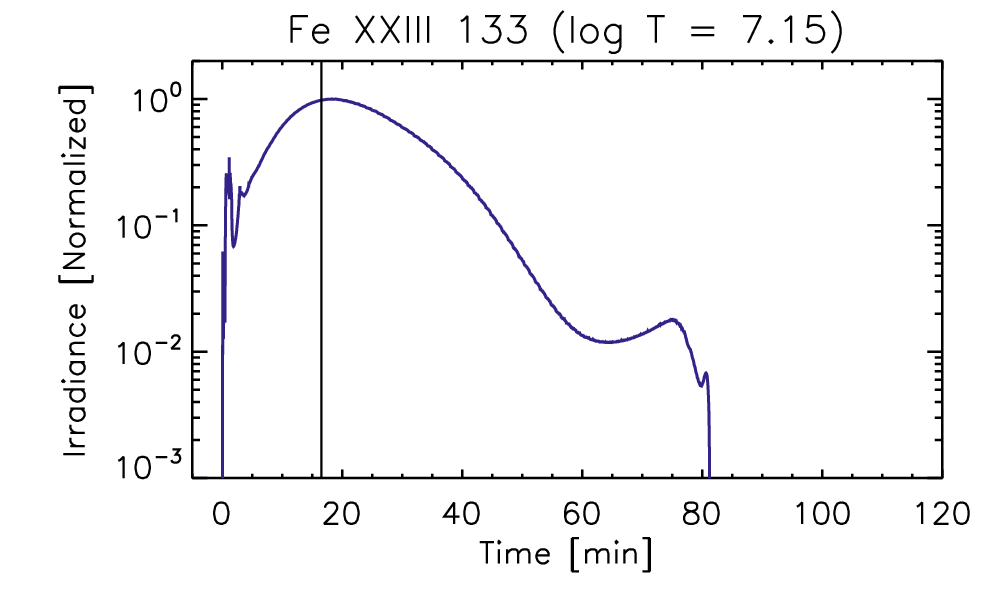}
    \includegraphics[width=0.48\textwidth]{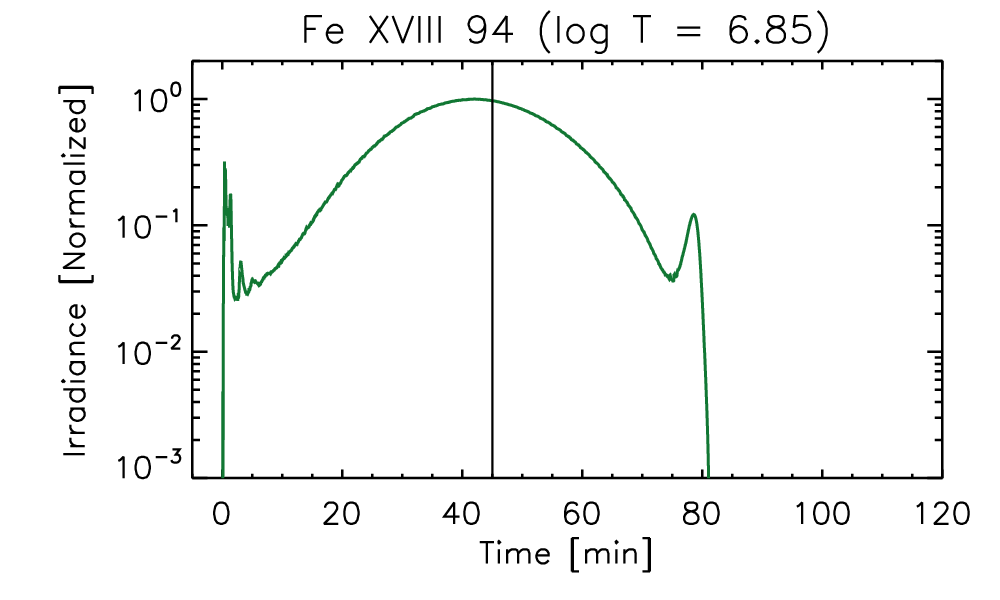}
    \includegraphics[width=0.48\textwidth]{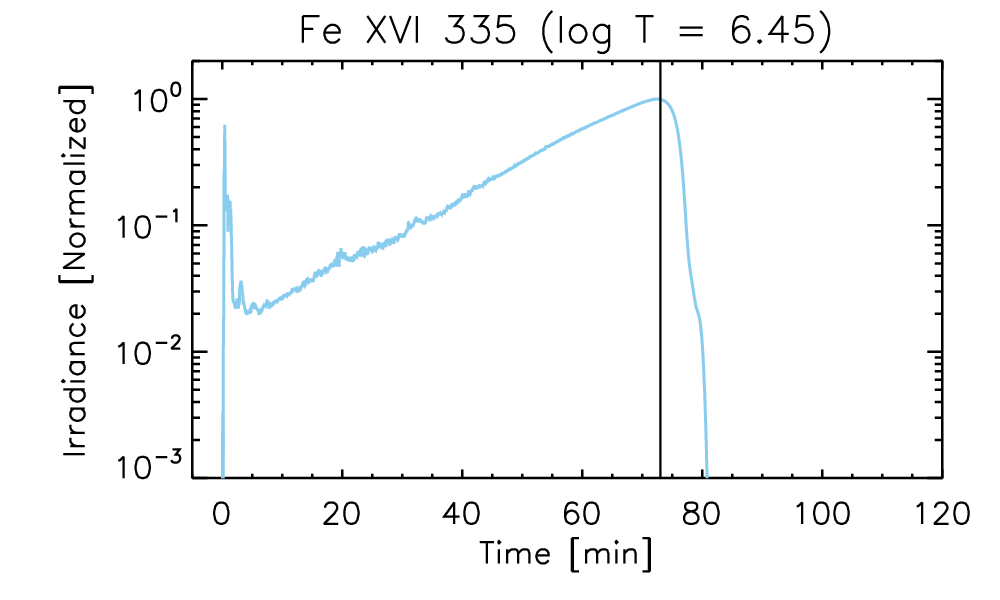}
    \includegraphics[width=0.48\textwidth]{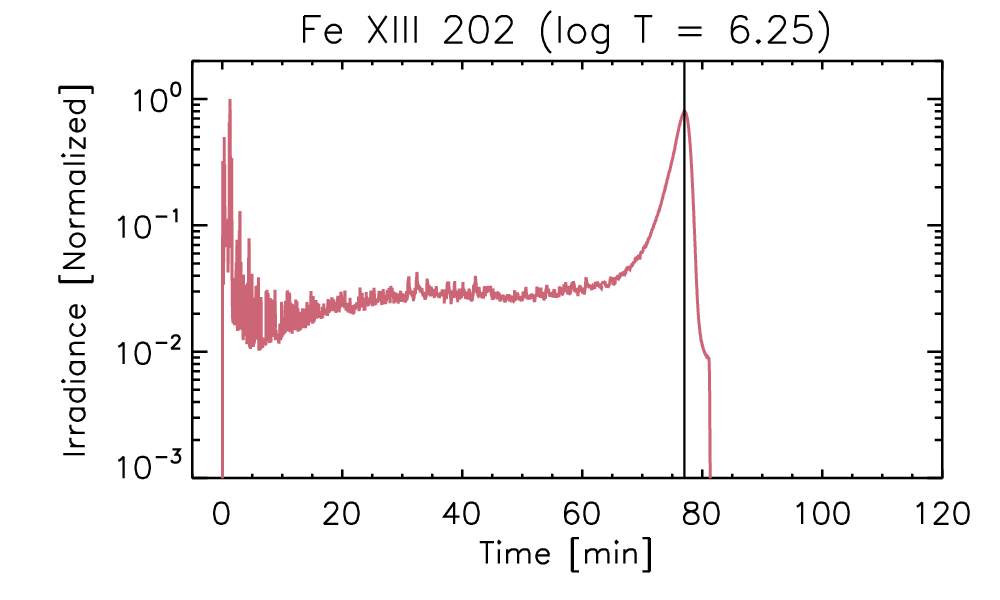}
    \includegraphics[width=0.48\textwidth]{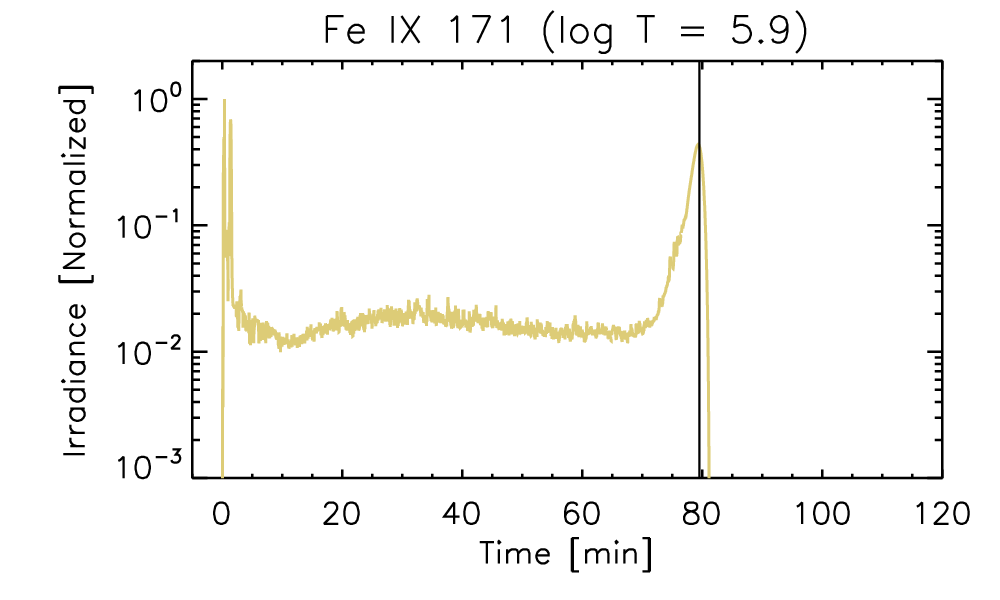}
    \includegraphics[width=0.48\textwidth]{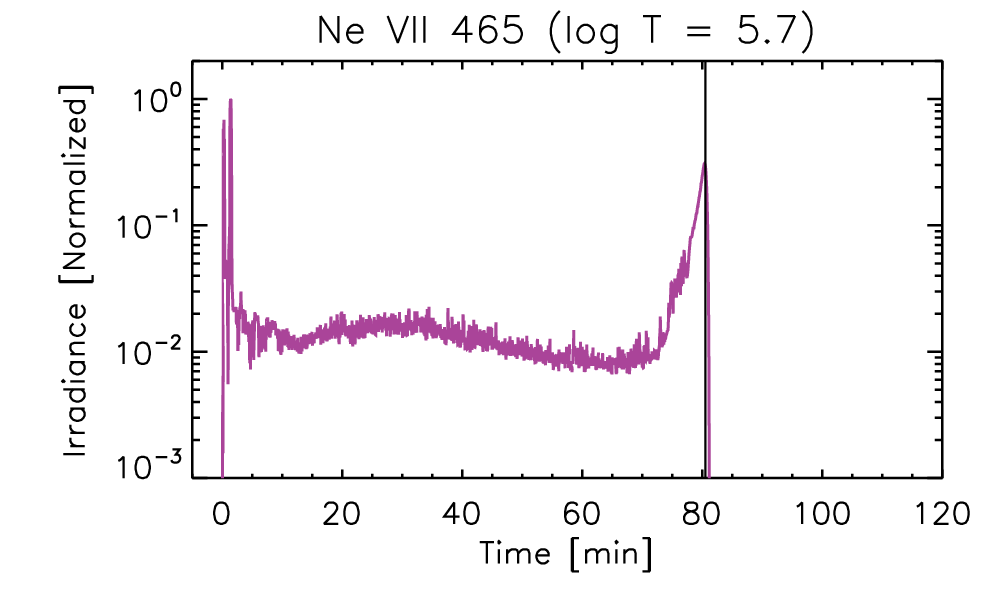}
    \caption{The cooling behavior of a 100 Mm loop with expanding cross section (factor of 11 from footpoint to apex) heated by a strong electron beam.  Similar to Figure \ref{fig:L100}.  The density remains constant through the radiative cooling phase, so the intensities track the change in ionization fractions.}
    \label{fig:exp}
\end{figure*}

Figure \ref{fig:exp} shows a similar set of plots for a loop with gradual area expansion across the length of the loop (by a factor of 11 from photosphere to apex).  Because of the area expansion, the flow of mass from the corona is significantly reduced (see detailed comparisons in \citealt{reep2022}), so that the cooling is dominated by radiation.  That is, for the scaling relation $T \propto n^{\delta}$, $\delta \rightarrow \infty$ and the cooling is driven by radiation.  The loop does not significantly drain until the onset of catastrophic cooling, so the coronal density remains constant, and energy is not lost through an enthalpy flux.  The cooling time is therefore longer than for a loop with uniform area.  

In Figure \ref{fig:exp}, the spectral lines evolve in two phases.  The first corresponds to the initial heating pulse, where the transition region is pushed downwards in height, raising the densities and causing intense brightenings.  This is particularly noticeable in the cool lines (\textit{e.g.} \ion{Fe}{9} 171 \AA).  The coronal density then rises from evaporation, and remains approximately constant through the entire cooling period.  As a result, the intensities of the lines change only due to changes in the ionization fraction.  Consider the evolution of the \ion{Fe}{18} 94 \AA\ case, where the intensity increases from about 20 to 40 minutes after the onset of heating, and then dims until about 70 minutes.  The coronal density is constant over this whole time period, so this change in intensity corresponds directly to a change in ionization fraction (recombining from \ion{Fe}{19} initially as it rises, then recombining into \ion{Fe}{17} as it fades).  This same behavior can be seen in all 6 lines.  Additionally, for the cooler lines, there is a strong component of emission from the transition region, that is roughly constant in density and ionization fraction after the initial heating pulse.  This can be seen in the roughly constant intensity in \textit{e.g.} \ion{Fe}{9} 171 \AA, until the coronal portion of the loop cools enough to begin forming this ion.

\begin{figure*}
    \centering
    \includegraphics[width=0.48\textwidth]{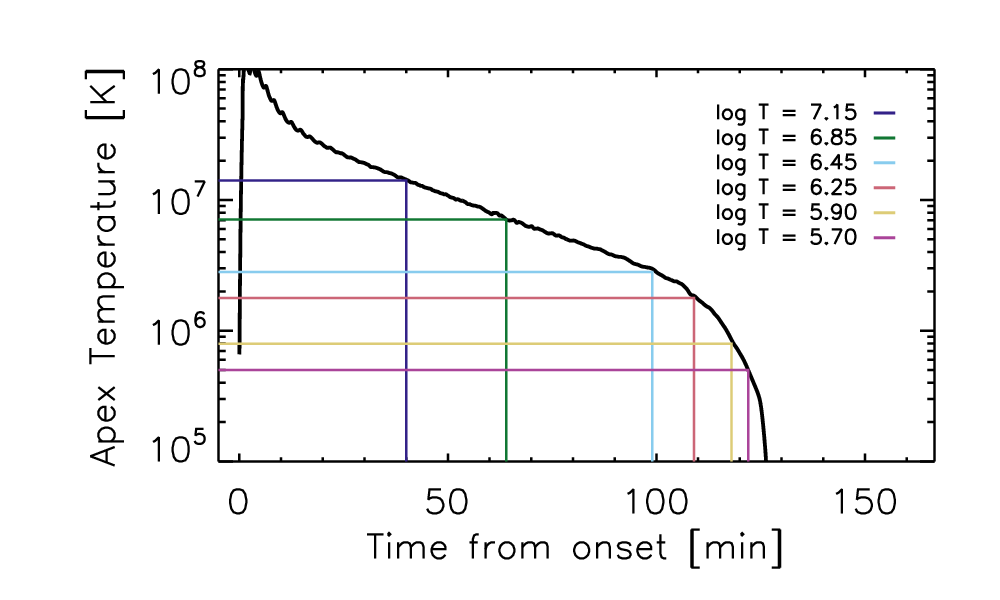}
    \includegraphics[width=0.48\textwidth]{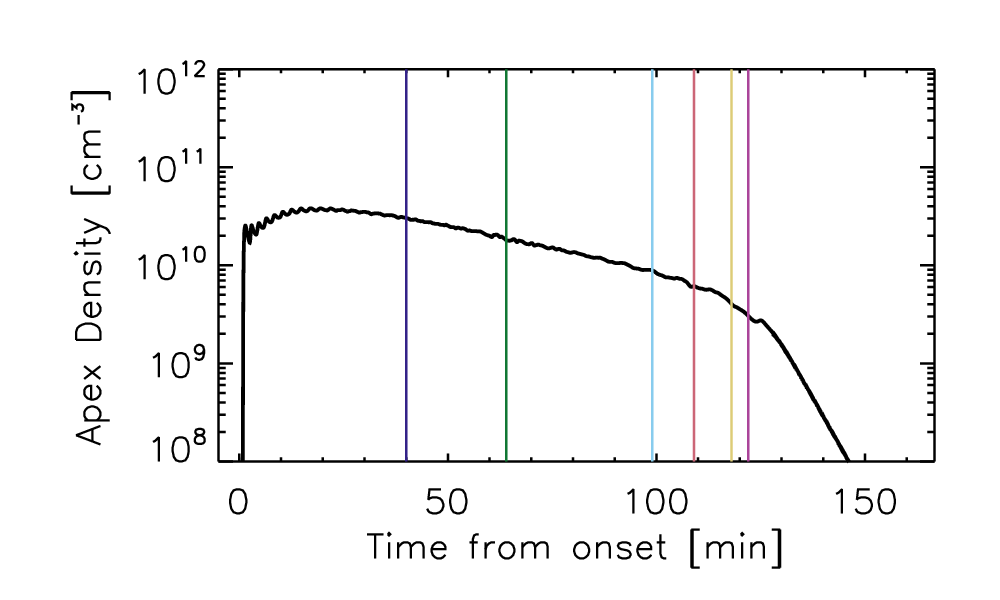}
    \includegraphics[width=0.48\textwidth]{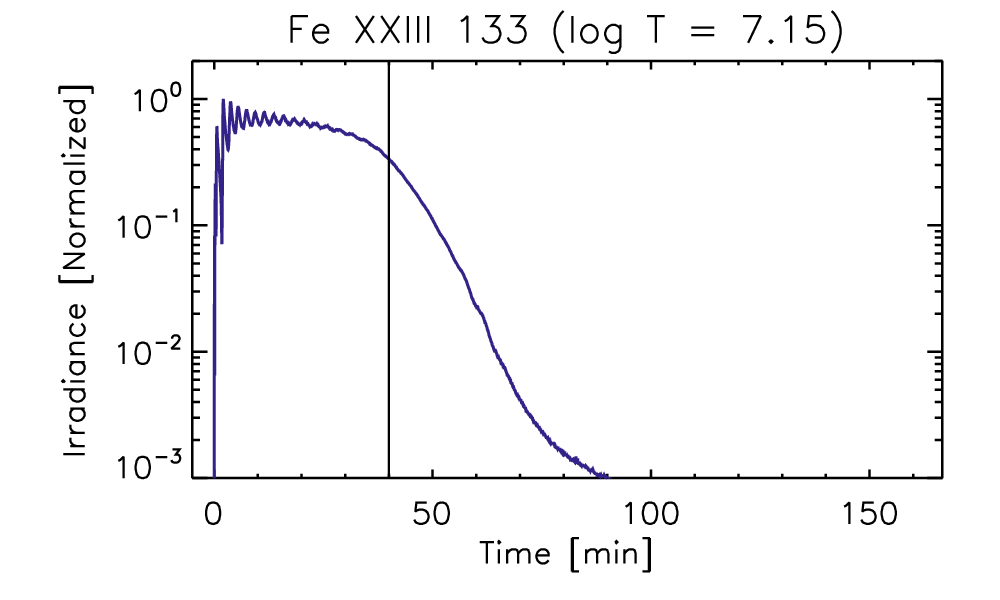}
    \includegraphics[width=0.48\textwidth]{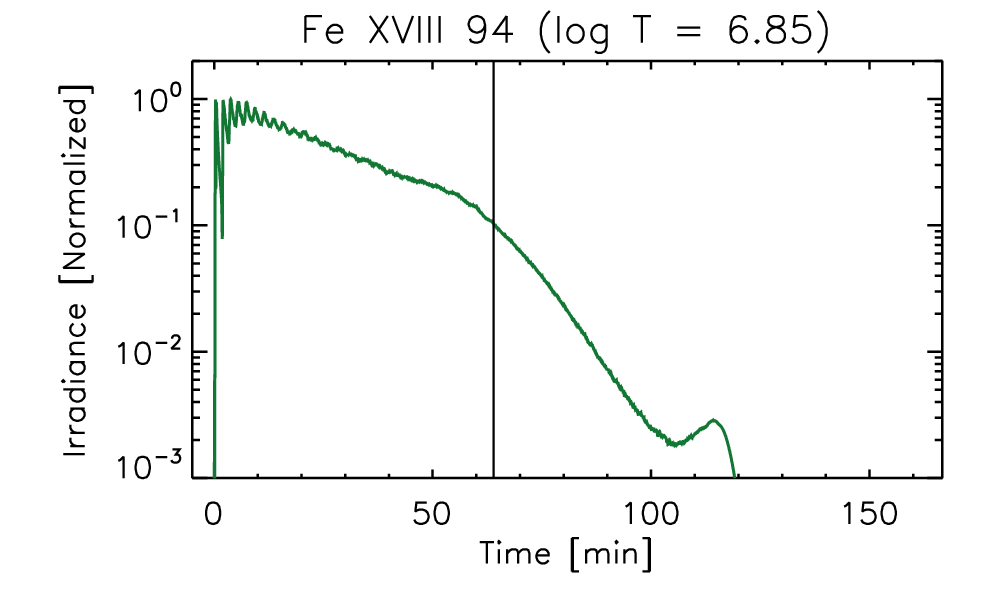}
    \includegraphics[width=0.48\textwidth]{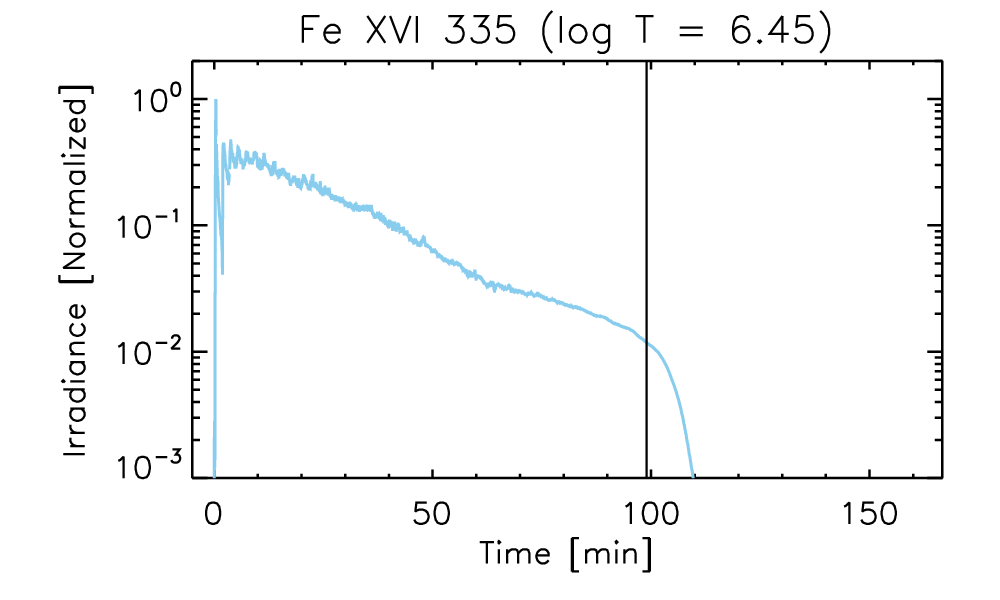}
    \includegraphics[width=0.48\textwidth]{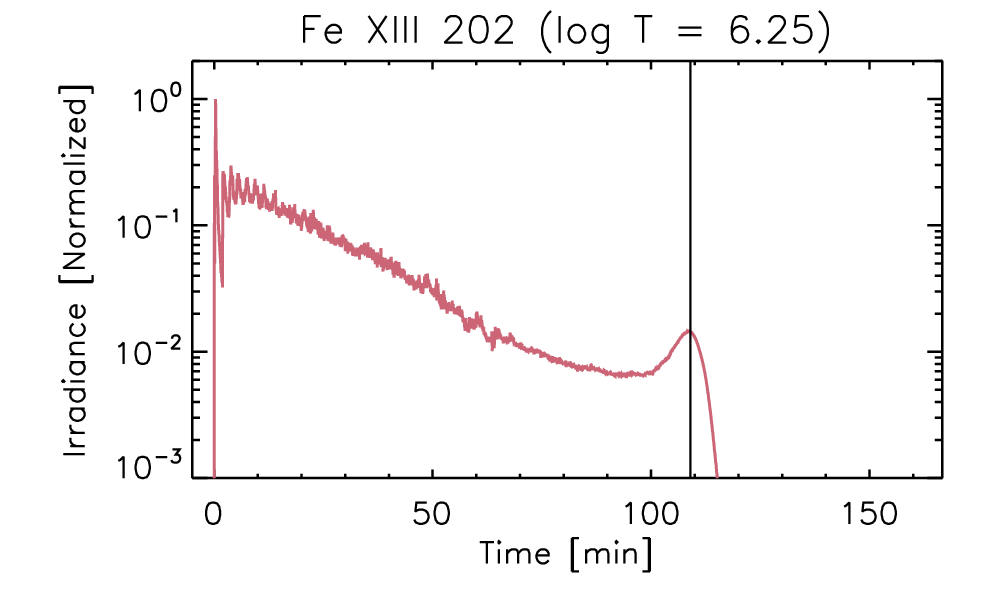}
    \includegraphics[width=0.48\textwidth]{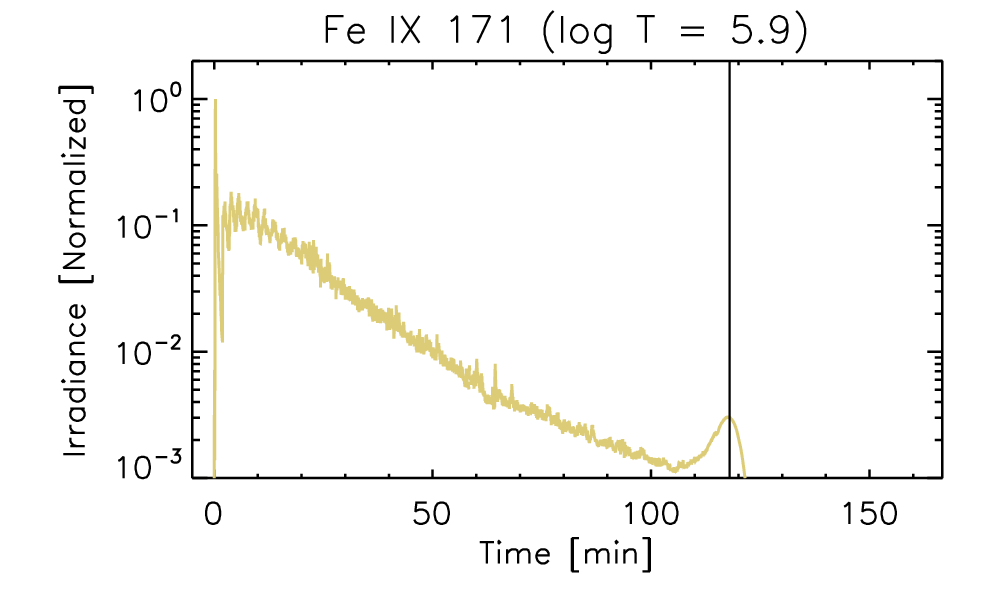}
    \includegraphics[width=0.48\textwidth]{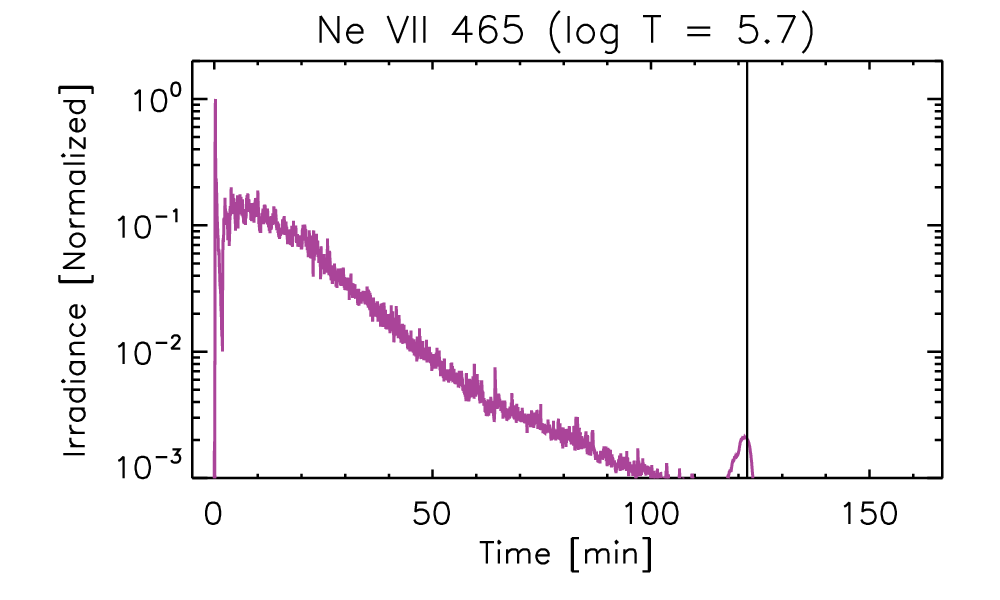}
    \caption{The cooling behavior of a uniform 100 Mm loop heated by a strong electron beam, with strong turbulent suppression of thermal conduction.  Similar to Figure \ref{fig:L100}.  The loop becomes much hotter and less dense than the laminar case, which causes the cooling and draining times to be significantly longer.}
    \label{fig:turb}
\end{figure*}

Next, we consider the case of a uniform loop with suppression of thermal conduction due to turbulence in Figure \ref{fig:turb}.  Following the formalism of \citet{bradshaw2019}, we set the turbulent mean-free path $\lambda_{T} = 10^{7}$ cm, which causes significant suppression of conduction.  We see from the apex temperature that the loop is heated to well over $10^{8}$ K as the energy is bottled up in the corona, gradually falling over time as radiation and flows carry away that energy.  Comparing to the case with Spitzer-H\"arm conduction, at the start of the cooling period, the temperature is significantly higher, and the density lower, so from Equation \ref{eqn:tauturb}, we can see that the cooling time should be significantly lengthened.  The loop slowly drains during the course of the cooling, additionally, which gradually reduces the radiative losses.

The time series of spectral lines in Figure \ref{fig:turb} show a rather different evolution than the previous cases.  There is still a spike in intensity corresponding to the onset of heating when the transition region is pushed downwards, and as before, it is much more noticeable in the cooler lines.  After that heating pulse, however, the intensities of the lines closely trace the evolution of the density.  The intensity of each rises as evaporation fills the loop, peaking at about 10 minutes after heating.  Then, as the density begins to drain, the intensity of each line begins to decrease.  There is an inflection point in each line, corresponding again to the time when the corona cools to the peak formation temperature of each ion.  So, for example, after about 40 minutes, the corona has cooled to $\log{T} = 7.15$, when \ion{Fe}{23} is most abundant, and there is an inflection point in the time series of the 133 \AA\ line where the intensity falls sharply as the corona recombines to cooler ions.  The same behavior can be seen in all six lines.

\begin{figure*}
    \centering
    \includegraphics[width=0.48\textwidth]{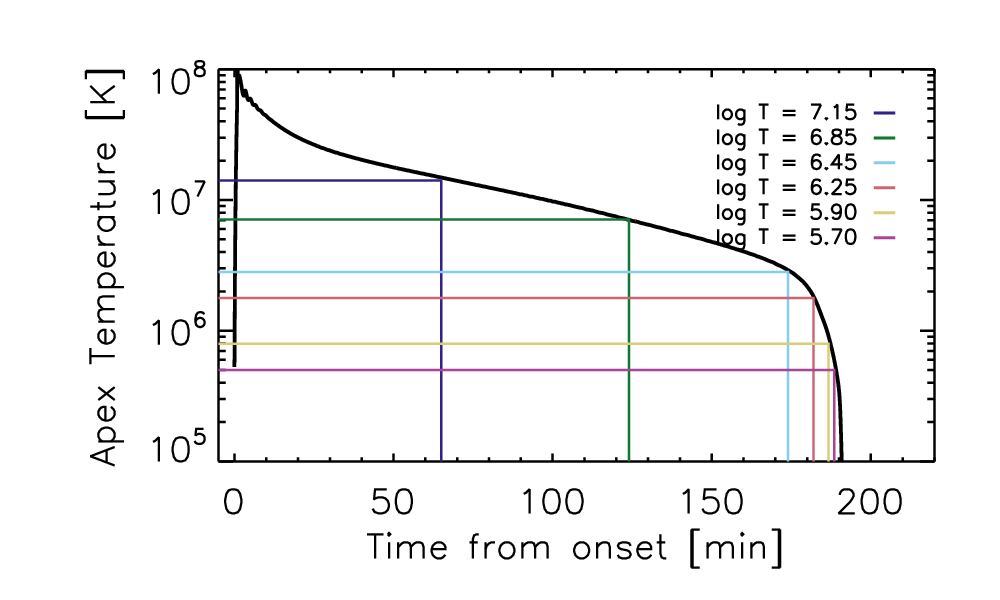}
    \includegraphics[width=0.48\textwidth]{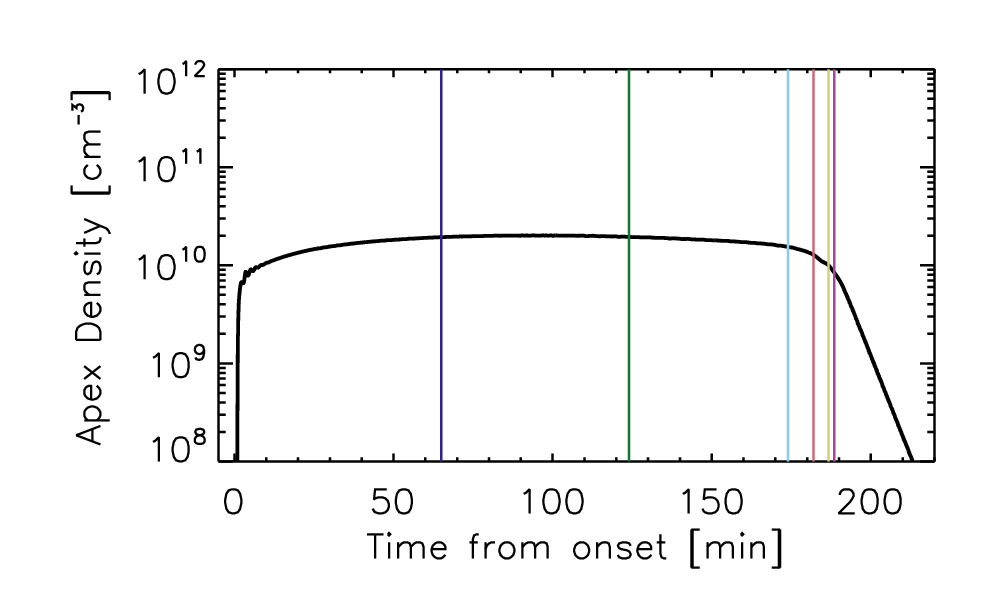}
    \includegraphics[width=0.48\textwidth]{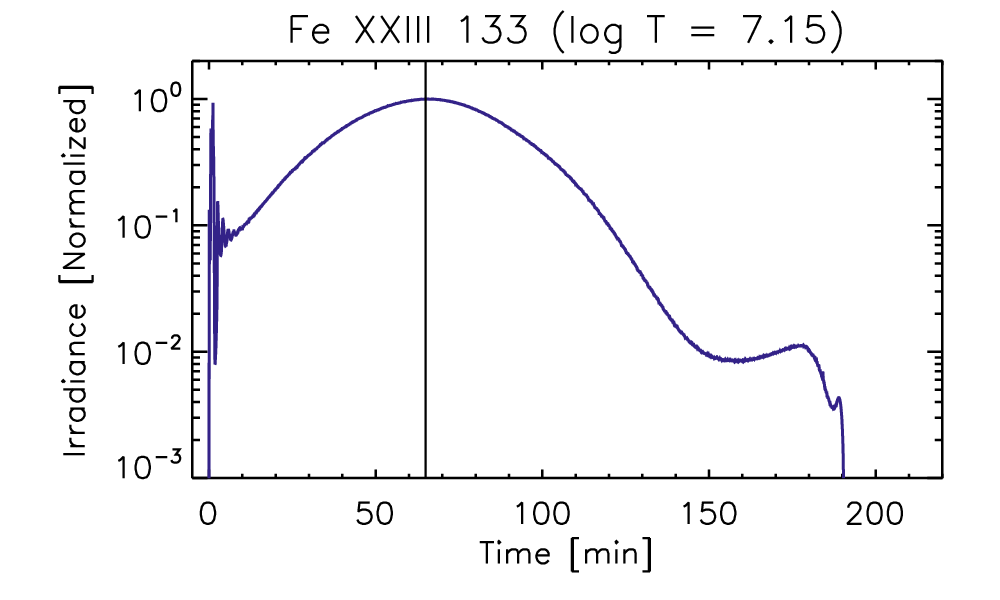}
    \includegraphics[width=0.48\textwidth]{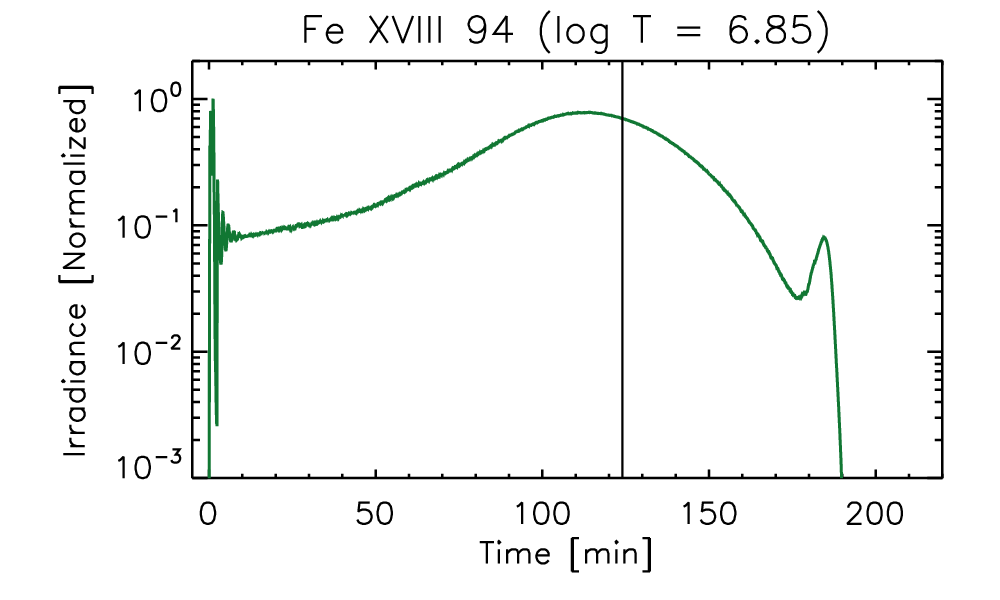}
    \includegraphics[width=0.48\textwidth]{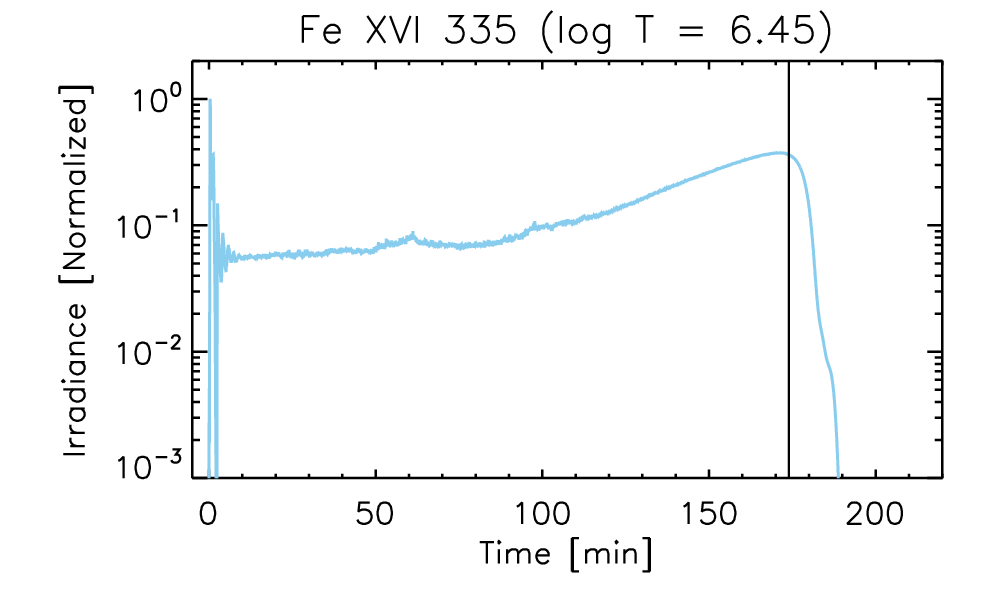}
    \includegraphics[width=0.48\textwidth]{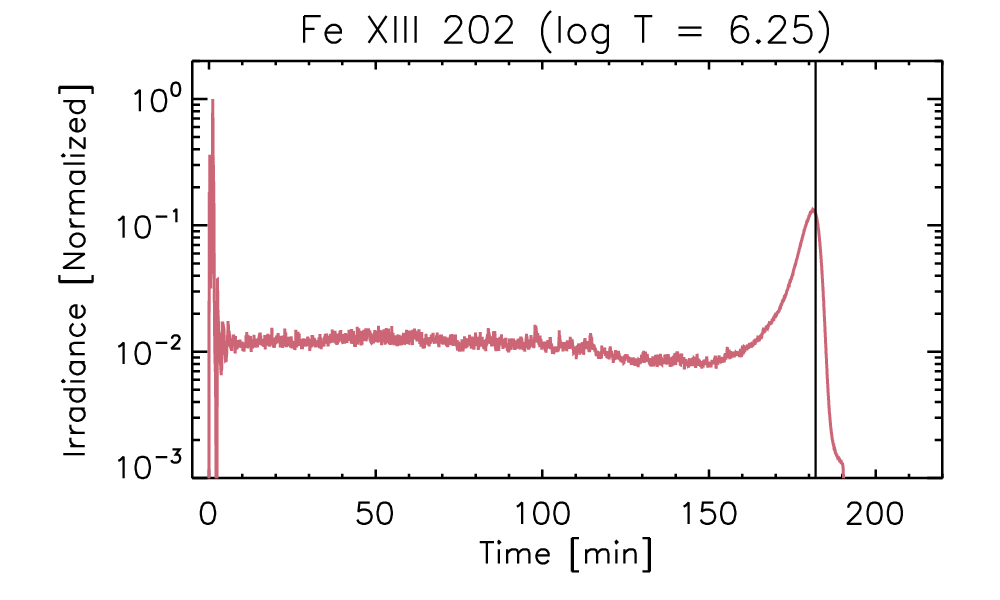}
    \includegraphics[width=0.48\textwidth]{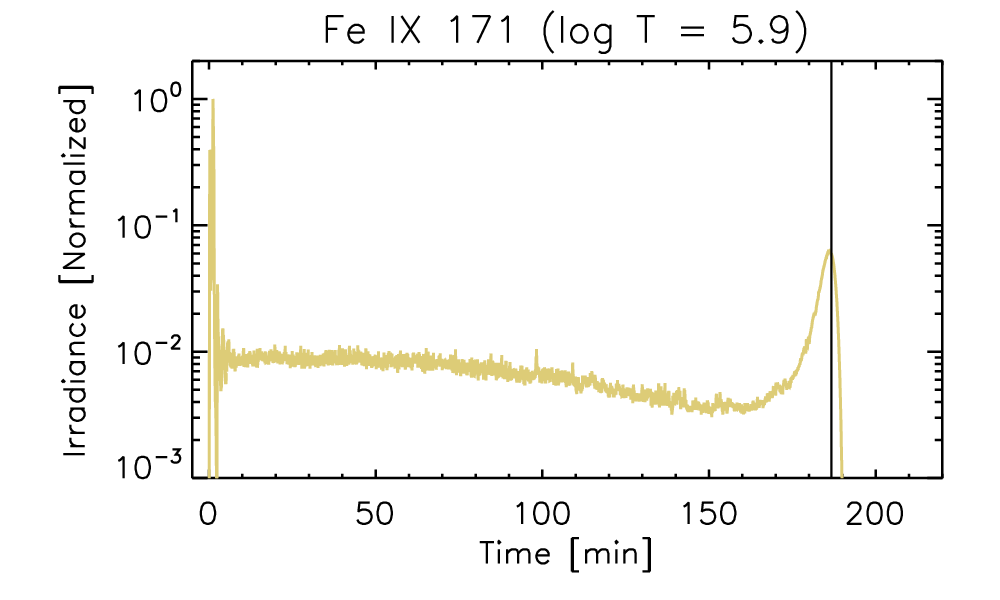}
    \includegraphics[width=0.48\textwidth]{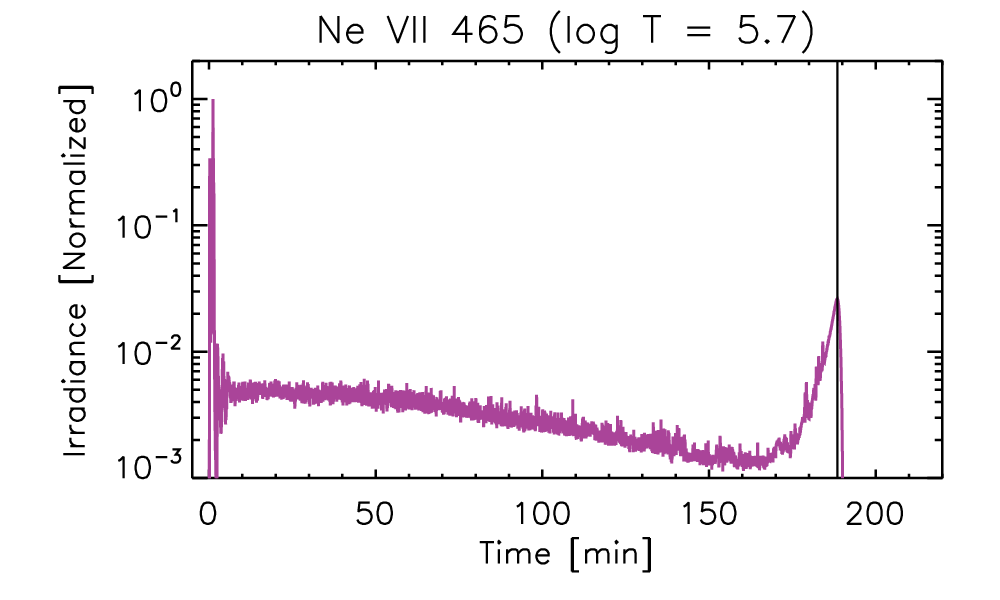}
    \caption{The cooling behavior of a 100 Mm loop with expanding cross section heated by a strong electron beam, with strong turbulent suppression of thermal conduction.  Similar to Figure \ref{fig:L100}.  Both thermal conduction and draining of the loop have been suppressed, so cooling proceeds only through radiation, and takes significantly longer than the previous cases.}
    \label{fig:te}
\end{figure*}

Finally, in Figure \ref{fig:te}, we consider the case of a loop with expanding cross section ($A_{max} = 11$) as well as turbulent suppression of thermal conduction ($\lambda_{T} = 10^{7}$ cm).  The behavior now shows an amalgamation of the previous two cases.  The coronal temperature rises to approximately $10^{8}$ K.  The coronal density rises with chromospheric evaporation to about $10^{10}$ cm$^{-3}$, but, due to the area expansion, does not drain significantly during the course of the cooling period and remains approximately constant until the onset of catastrophic cooling.  Since cooling due to conduction and enthalpy flux have both been suppressed in this case, radiation is the only effective cooling mechanism, and this takes significantly longer than in any previous case.   

The evolution of the spectral lines in Figure \ref{fig:te} shows behavior similar to those in Figure \ref{fig:exp}.  There is a sharp spike in intensity due to the initial heating pulse, causing the transition region to be pushed to lower heights and higher densities.  Then, the corona is heated and filled, but is too hot to strongly emit in any of the six lines.  In this case, all six lines initially form in the transition region as the corona is far too hot for any of the ions to be abundant.  As the corona slowly cools, however, each ion begins to form as hotter ions recombine.  The coronal density remains constant for the cooling period, so once again the change in intensity closely corresponds to the change in ionization fractions.  As a result, each has a roughly constant component emanating from the transition region that produces a weak but constant intensity.  The coronal component dominates when the corona reaches the peak formation temperature of each ion, and then rapidly falls as the corona cools further.  

\begin{figure*}
    \centering
    \includegraphics[width=0.48\textwidth]{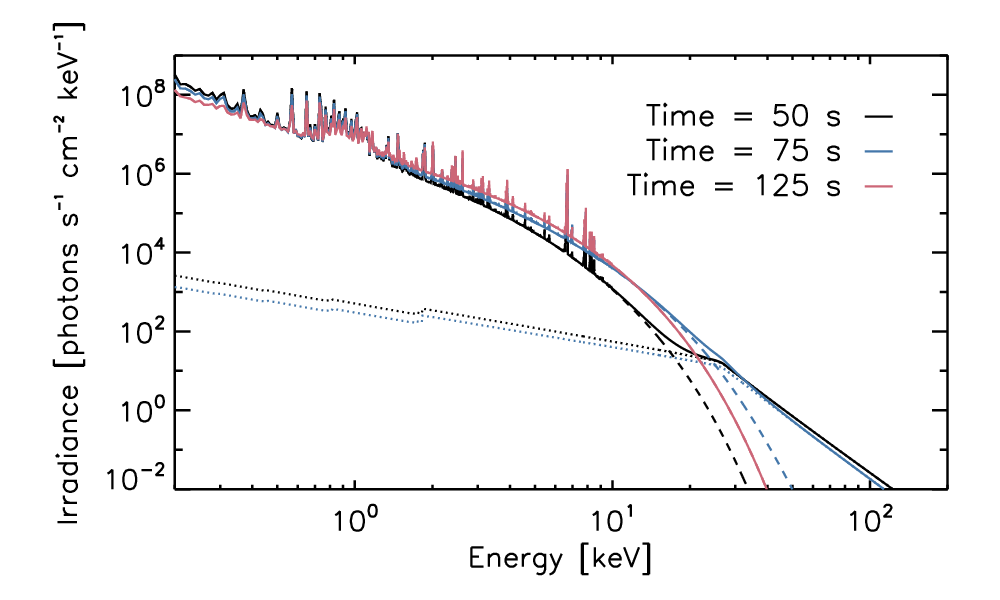}
    \includegraphics[width=0.48\textwidth]{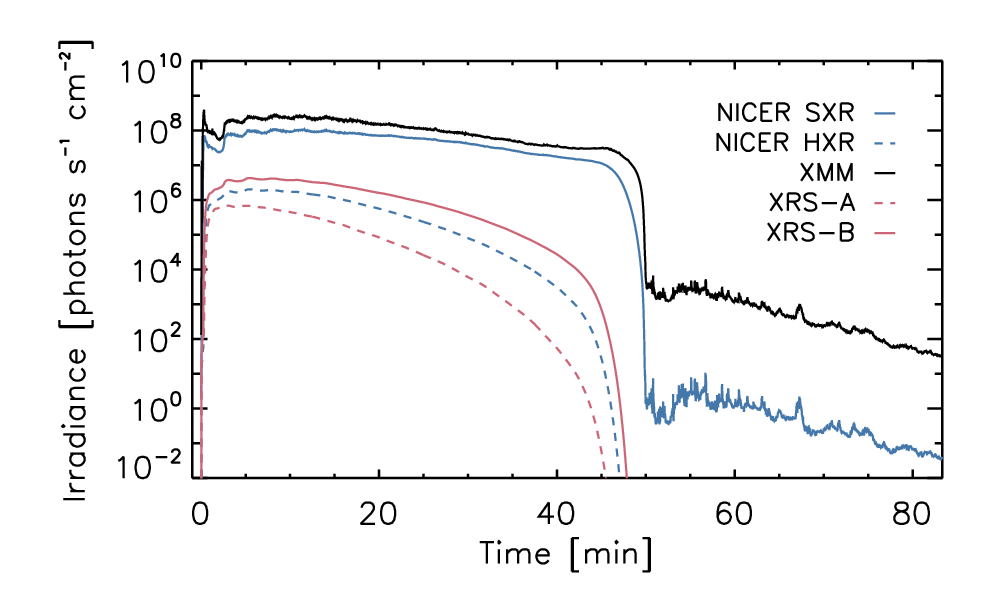}
    \includegraphics[width=0.48\textwidth]{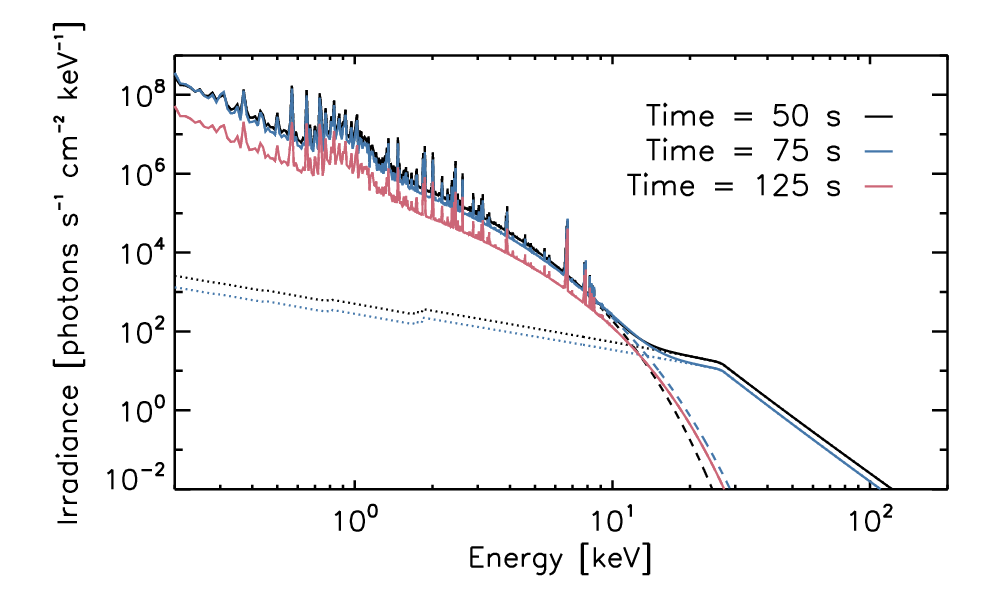}
    \includegraphics[width=0.48\textwidth]{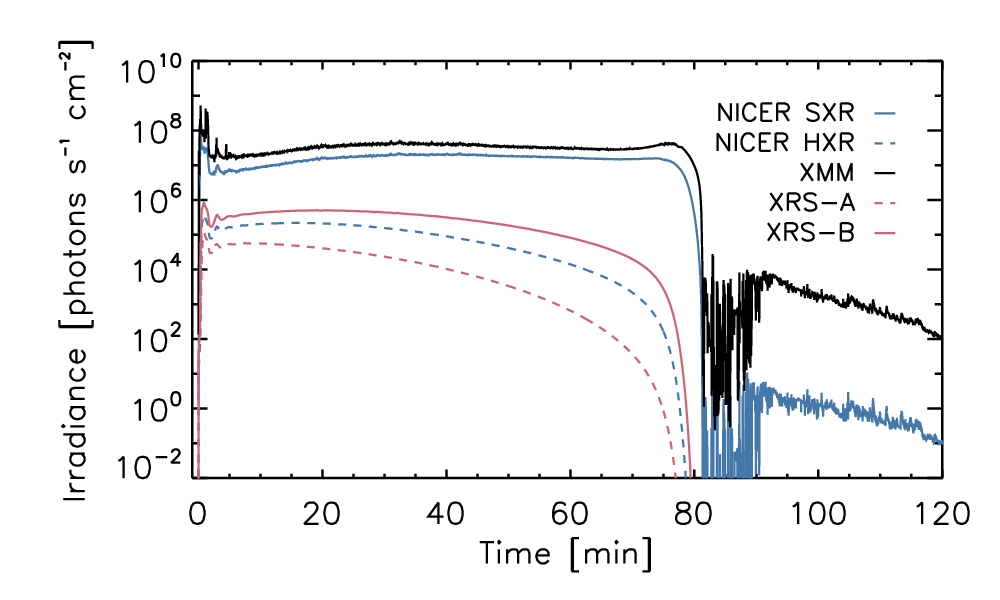}
    \includegraphics[width=0.48\textwidth]{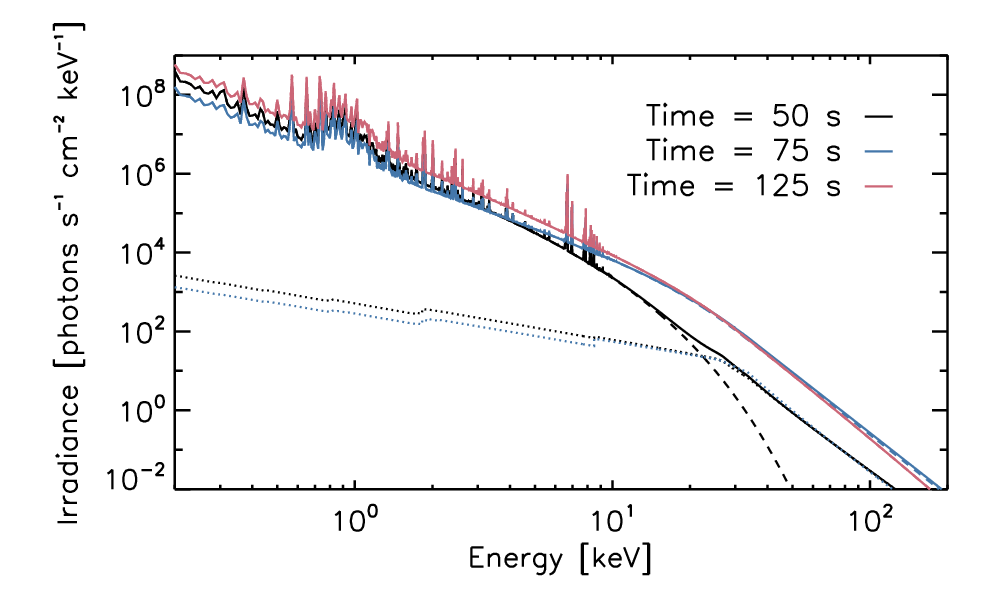}
    \includegraphics[width=0.48\textwidth]{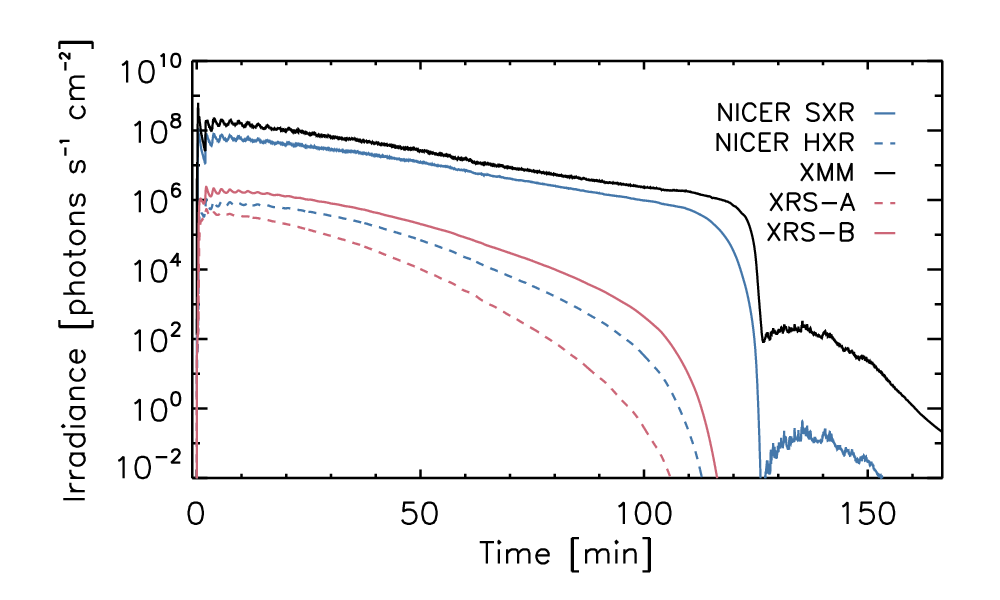}
    \includegraphics[width=0.48\textwidth]{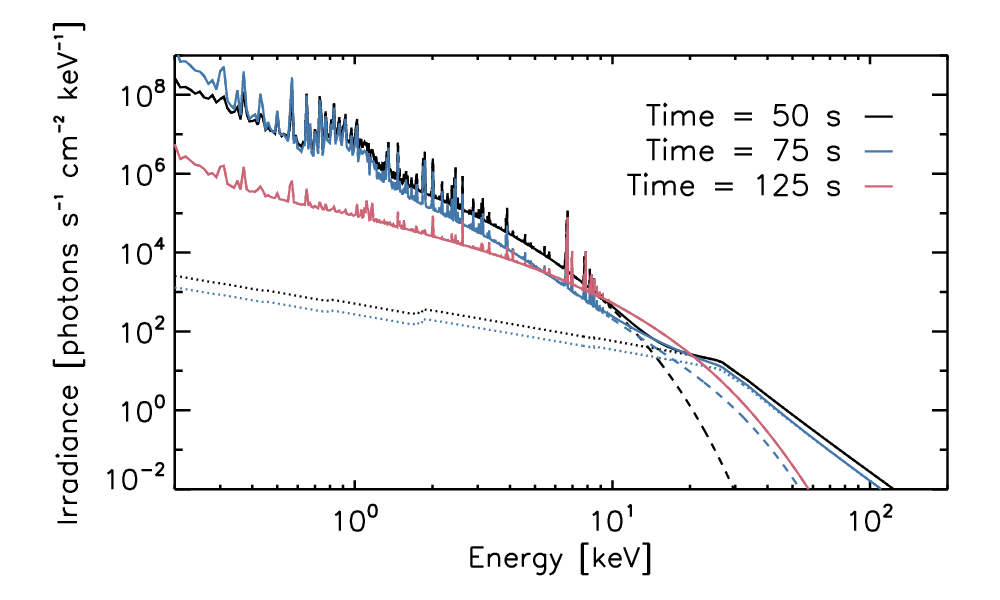}
    \includegraphics[width=0.48\textwidth]{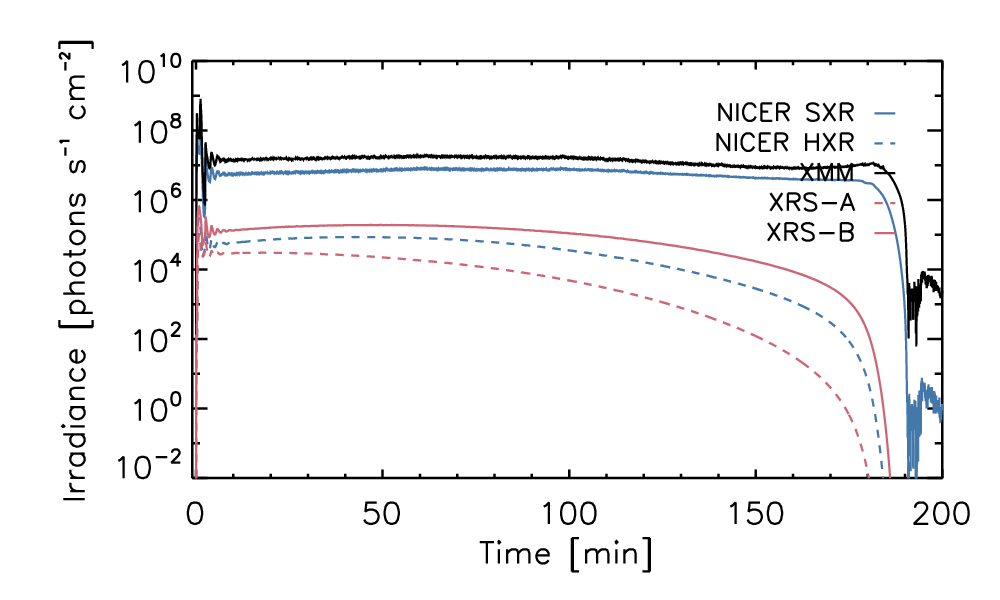}
    \caption{The evolution of X-ray emission in the four loop simulations described in Figures \ref{fig:L100} through \ref{fig:te}, from top to bottom, respectively.  The left-hand column shows the X-ray spectrum at the time of peak heating (50 s), as the heating decays (75 s), and at the onset of the cooling phase (125 s).  The right-hand column shows the irradiance time series in the energy ranges observed by NICER, Newton/XMM, or GOES/XRS (assuming a distance of 1 AU). }
    \label{fig:xrays}
\end{figure*}

We briefly turn our attention to the X-rays, which are used to classify solar flares and used for diagnosing particle acceleration.  Figure \ref{fig:xrays} shows synthetic X-ray emission from each of the four simulations in Figures \ref{fig:L100} to \ref{fig:te}, from top to bottom respectively.  The plots in the left column show the X-ray spectra at three different times corresponding to the peak of the heating event (50 s), decay of the heating event (75 s), and near the start of the cooling period (125 s).  The total spectrum (solid) is calculated by combining the non-thermal emission (dotted) from the electron beam with the thermal line and continuum emission (dashed).  The spectra were calculated at 0.01 keV resolution.  The emission is dominated by thermal bremsstrahlung at low energies, while the non-thermal bremsstrahlung dominates at high energies during the heating period.  We then compare time series using energy ranges from three X-ray instruments, the Neutron Interior Composition Explorer (NICER; \citealt{gendreau2016}), X-ray Multi-Mirror Mission (XMM-Newton; \citealt{jansen2001}), and GOES/XRS.  The right column shows the time series of the X-rays in physical units corresponding to five energy ranges observed by NICER SXR (0.2 to 2 keV), NICER HXR (2 to 12 keV), XMM (0.2 to 20 keV), GOES XRS/A (3 to 24 keV), and GOES XRS/B (1.5 to 12 keV).  In each case, the softer energy ranges for XMM and NICER SXR remain bright for longer than for the others.  The high energy emission, dominated by thermal bremsstrahlung, decays rapidly as the temperature of the loop decreases.  That is, time series of higher energy X-ray bands should be shorter in duration than lower energy bands because of the strong dependence of bremsstrahlung on temperature. 

\begin{figure*}
    \centering
    \includegraphics[width=0.95\textwidth]{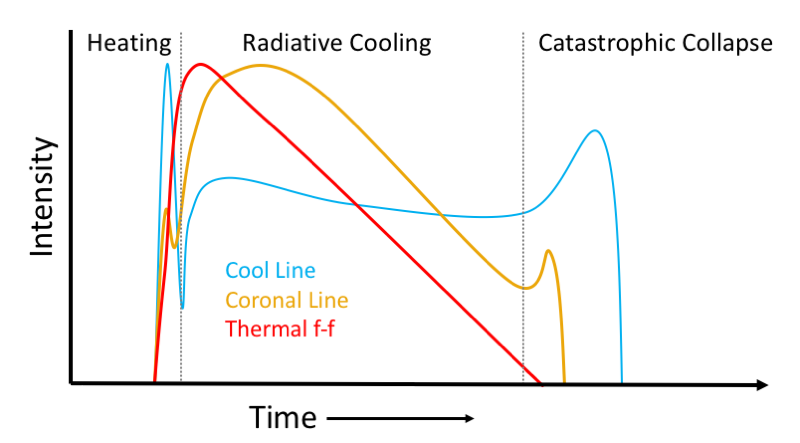}
    \caption{A cartoon description of the phases of evolution of emission in a coronal loop.  There are generally three phases: a heating phase, a radiative cooling phase, and catastrophic collapse.  The example light curves show a cool line (blue), forming near the lower transition region or upper chromosphere, a coronal line (gold), forming in the corona and upper transition region, and thermal free-free emission (red), forming in the corona.  The evolution depends strongly on the height at which emission forms.}
    \label{fig:cartoon}
\end{figure*}

The time evolution of emission, and thus the duration, depends on the height at which the emission forms, the relative importance of the three cooling mechanisms, and the details of the heating.  We summarize the cooling evolution and resultant emission with a cartoon description in Figure \ref{fig:cartoon}, including a cool spectral line, a coronal line, and thermal free-free emission.  There are generally three phases of evolution.  In the first phase, the onset of heating causes a rise in temperature in the corona through the upper chromosphere.  The transition region is pushed to lower heights, and chromospheric evaporation begins.  Lines forming in the transition region or upper chromosphere, as well as non-thermal emission from accelerated particles, spikes sharply during this phase.  Hotter emission begins to rise with the temperature, and grows slowly as the coronal density increases (the Neupert effect).  During the second phase, the onset of cooling, evaporation slows and the coronal density reaches its highest value.  The plasma initially cools from thermal conduction, but transitions to a radiative cooling phase.  If the cross-sectional area of the loop is close to uniform, it will slowly drain during this period, while large area expansions will inhibit draining.  Thermal bremsstrahlung peaks near the onset of cooling, and rapidly falls off as the plasma cools.  Hot lines grow in intensity as the coronal plasma approaches their peak formation temperature, and then begin to fall off as the ions recombine to lower ionization states.  Cooler lines, forming in the transition region at this time or below, remain approximately constant in intensity.  Finally, at the onset of catastrophic collapse, the temperature plummets rapidly in the corona.  The hot lines, which may have some residual emission, may have a small spike here before disappearing.  As the coronal portion of the loop cools, cooler ions, previously only present in the transition region, begin to form in the corona, and there is a large spike in emission at this point.  As the temperature continues to plummet and material rapidly drains from the loop, all emission fades away rapidly.  

\section{Parameter Survey}

\begin{figure*}
    \centering
    \includegraphics[width=0.48\textwidth]{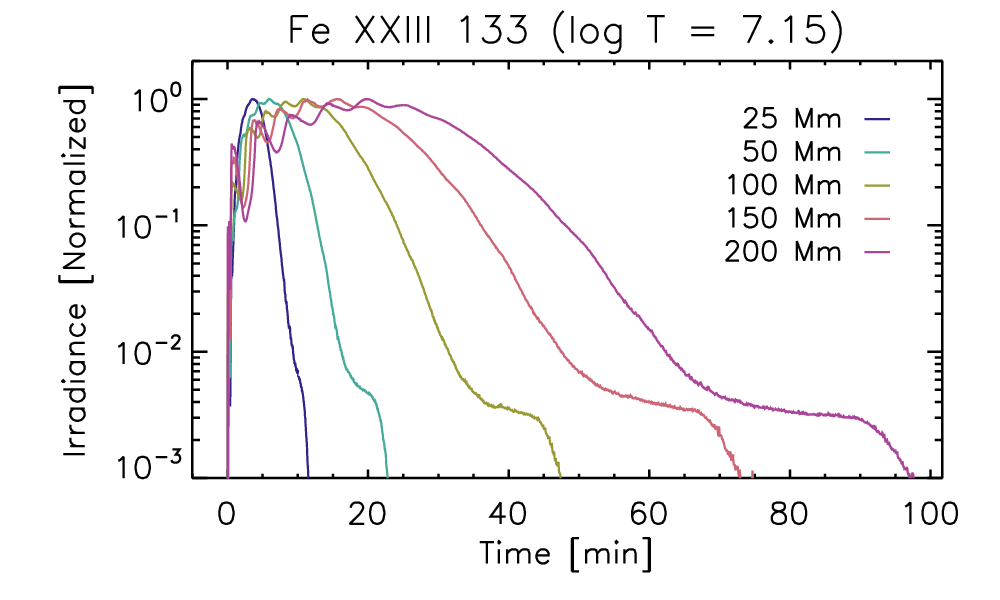}
    \includegraphics[width=0.48\textwidth]{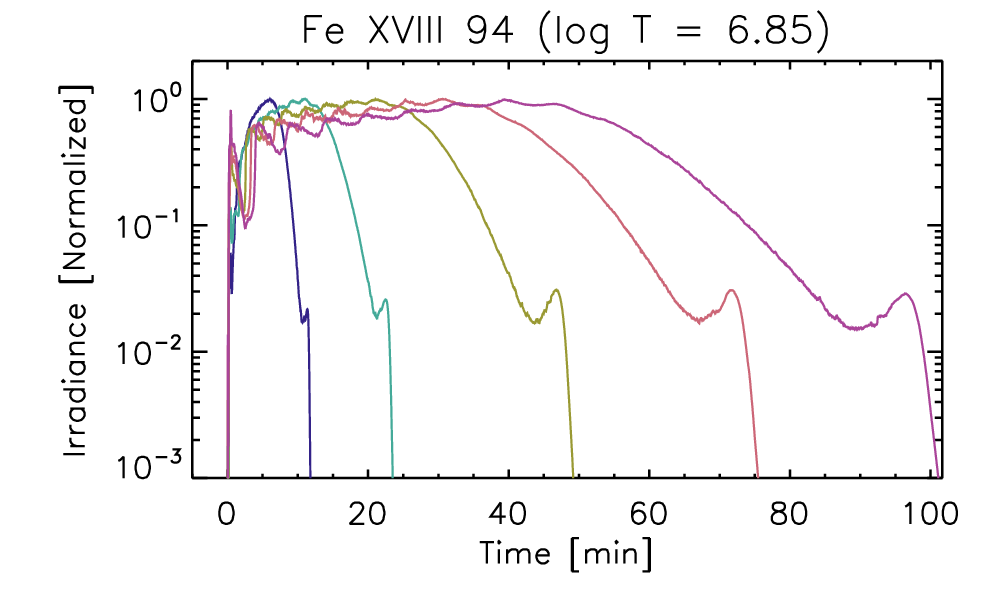}
    \includegraphics[width=0.48\textwidth]{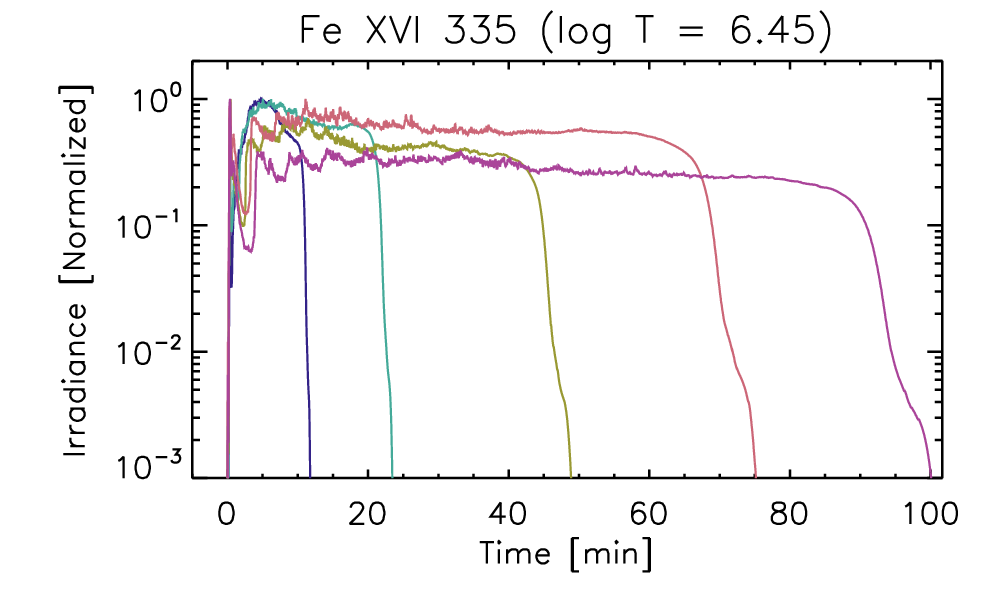}
    \includegraphics[width=0.48\textwidth]{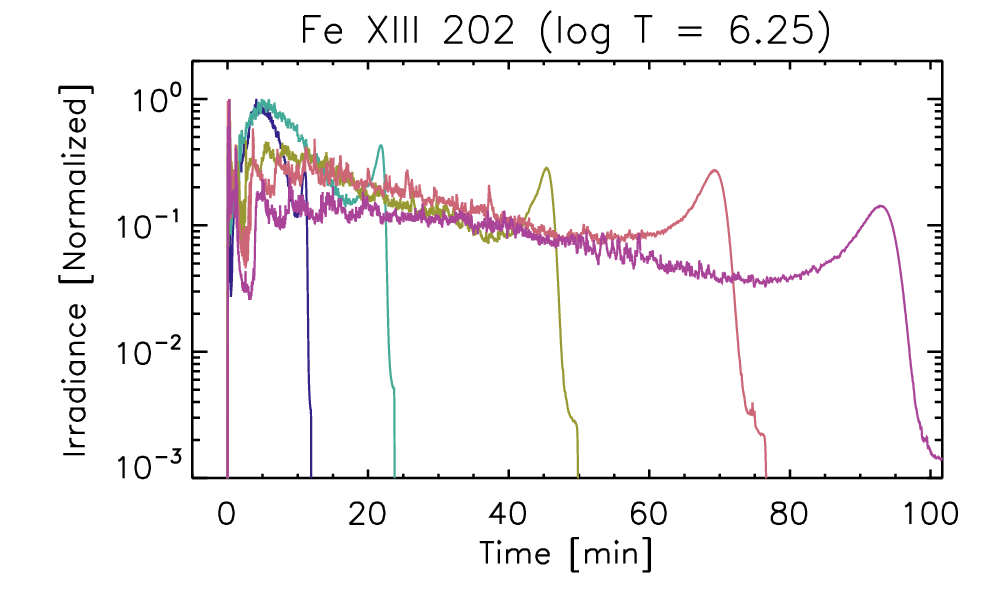}
    \includegraphics[width=0.48\textwidth]{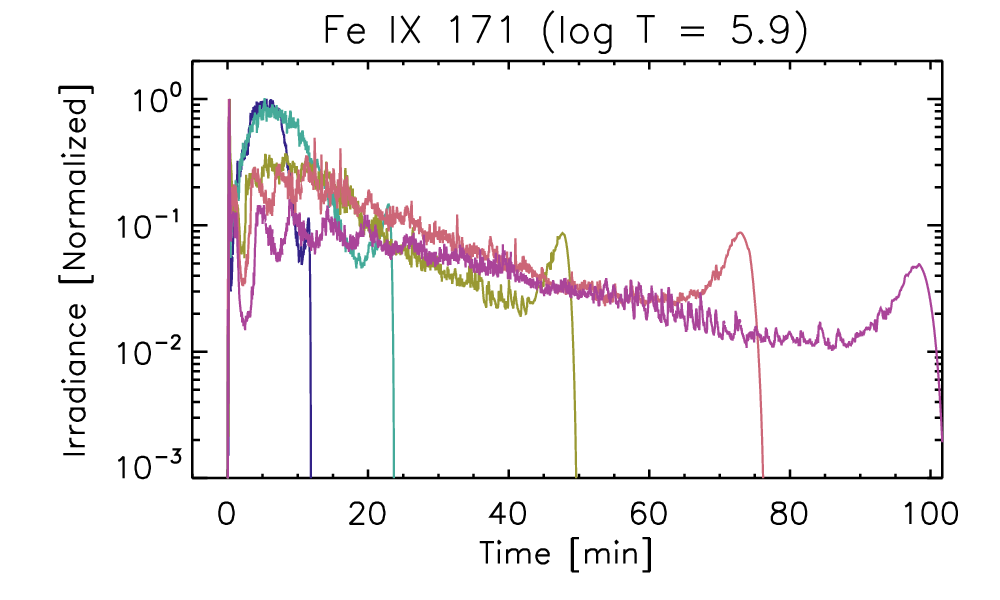}
    \includegraphics[width=0.48\textwidth]{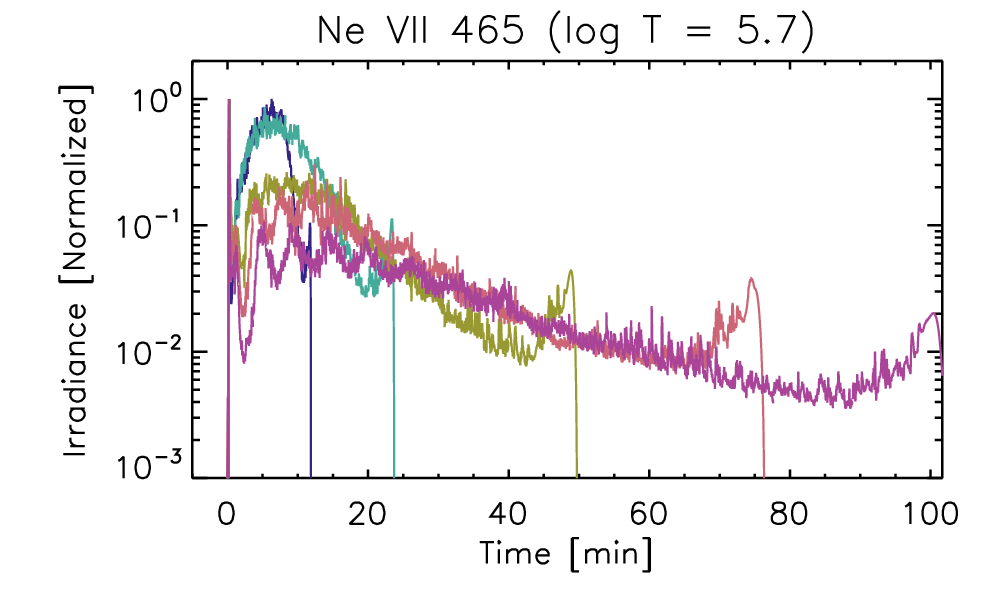}
    \caption{Irradiance time series for uniform, laminar loops of total lengths between 25 and 200 Mm.  The duration depends strongly on loop length as well as the formation temperature of each spectral line.  The intensity spike during the heating period is relatively stronger in longer loops.}
    \label{fig:length}
\end{figure*}

We briefly turn to examine how various parameters can impact the duration of light curves.  In Figure \ref{fig:length}, we show a comparison of the time series for the same six spectral lines as before with five loop lengths, [25, 50, 100, 150, 200] Mm.  We use a uniform cross-sectional area, and Spitzer-H\"arm conduction in these cases.  As derived by \citet{cargill1995}, the total cooling time of a loop scales as $L^{5/6}$, so the duration of each time series grows with loop length.  As pointed out in Figure \ref{fig:L100} and accompanying text, the duration also differs for each spectral line, with hotter lines fading away sooner.  Additionally, in longer loops, the density increase from evaporation is relatively smaller than for shorter loops, so the intensity spike during the heating period is more prominent in longer loops.

\begin{figure*}
    \centering
    \includegraphics[width=0.48\textwidth]{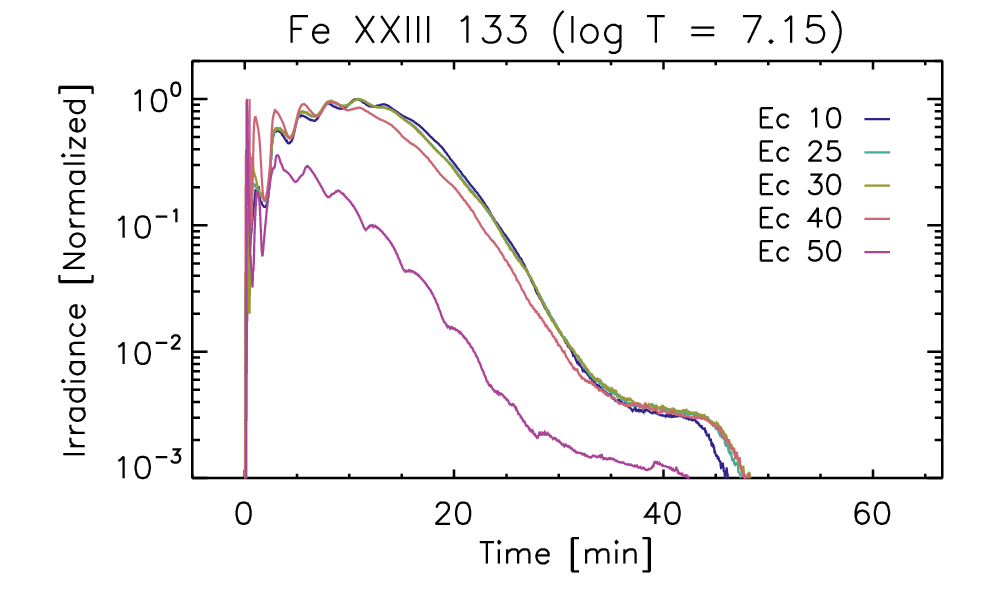}
    \includegraphics[width=0.48\textwidth]{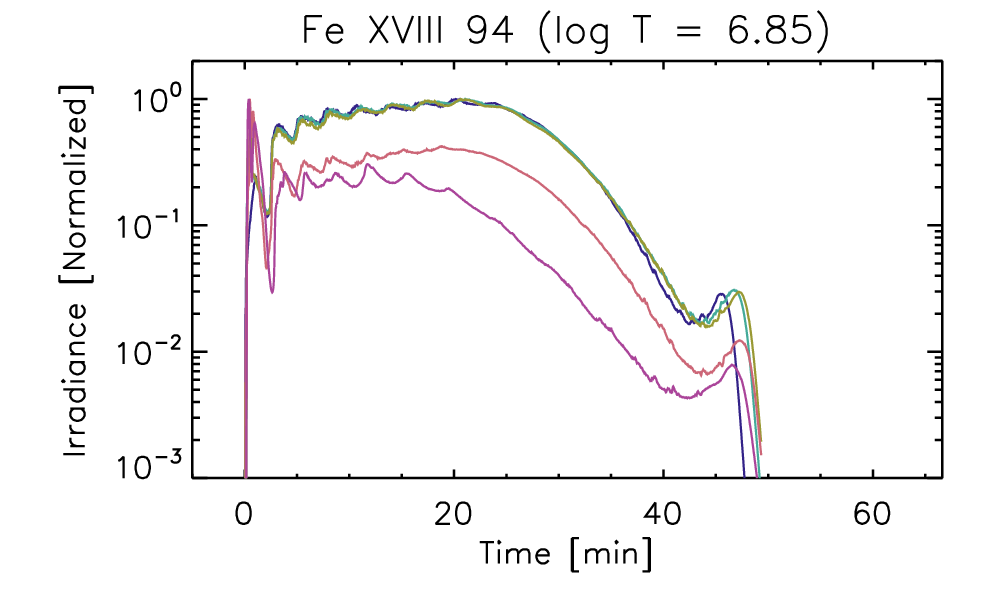}
    \includegraphics[width=0.48\textwidth]{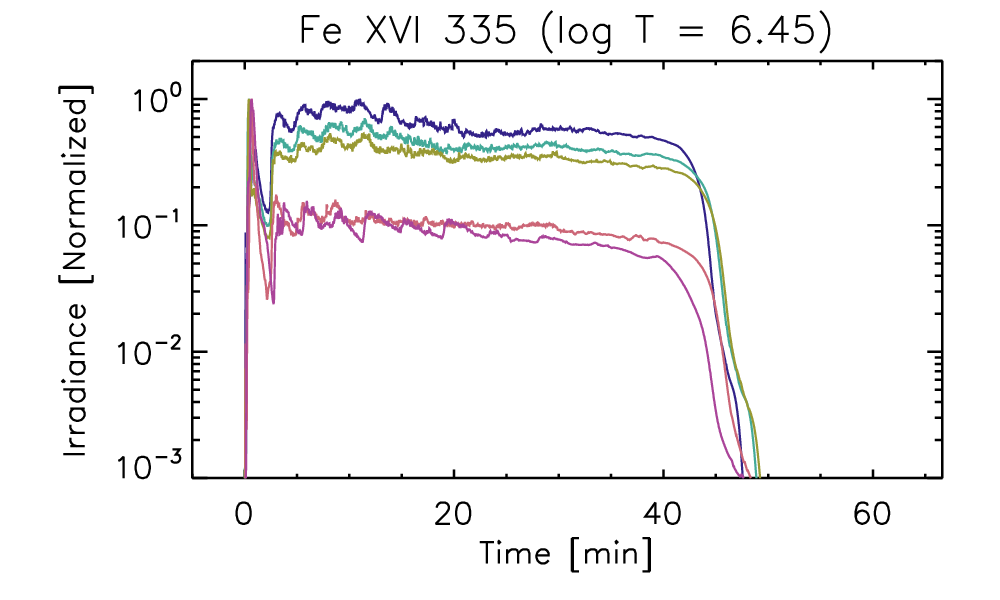}
    \includegraphics[width=0.48\textwidth]{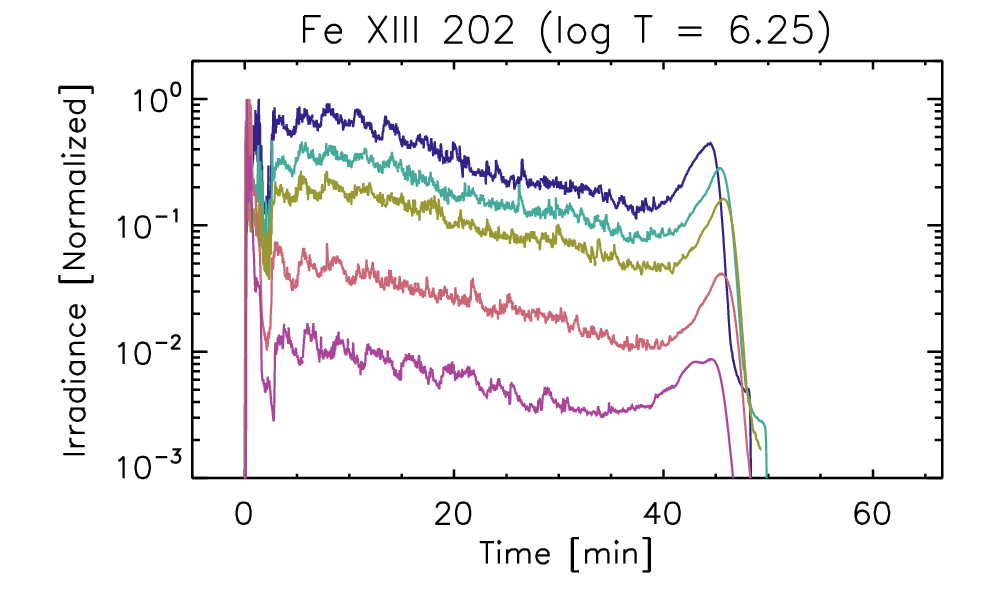}
    \includegraphics[width=0.48\textwidth]{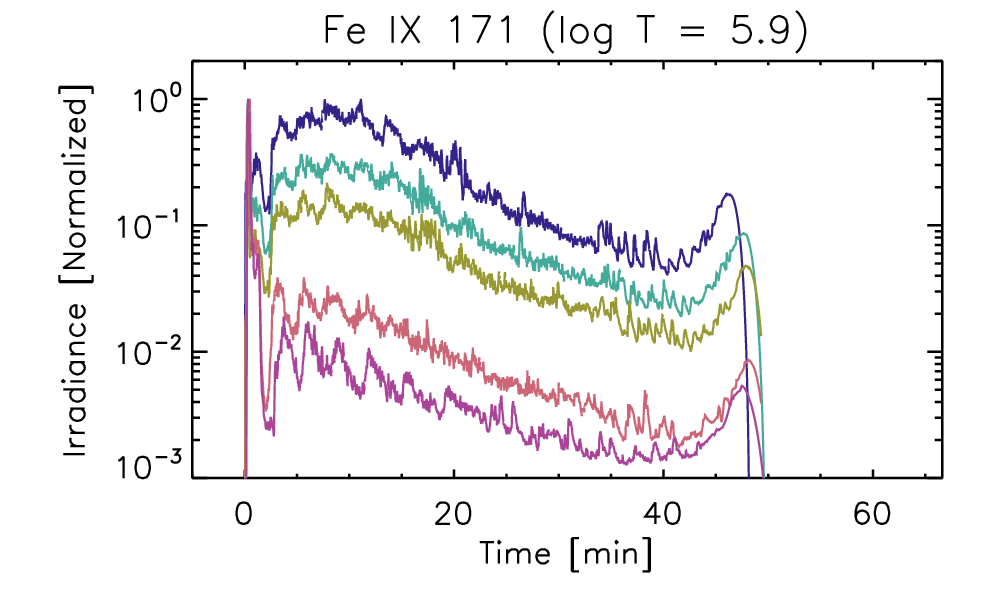}
    \includegraphics[width=0.48\textwidth]{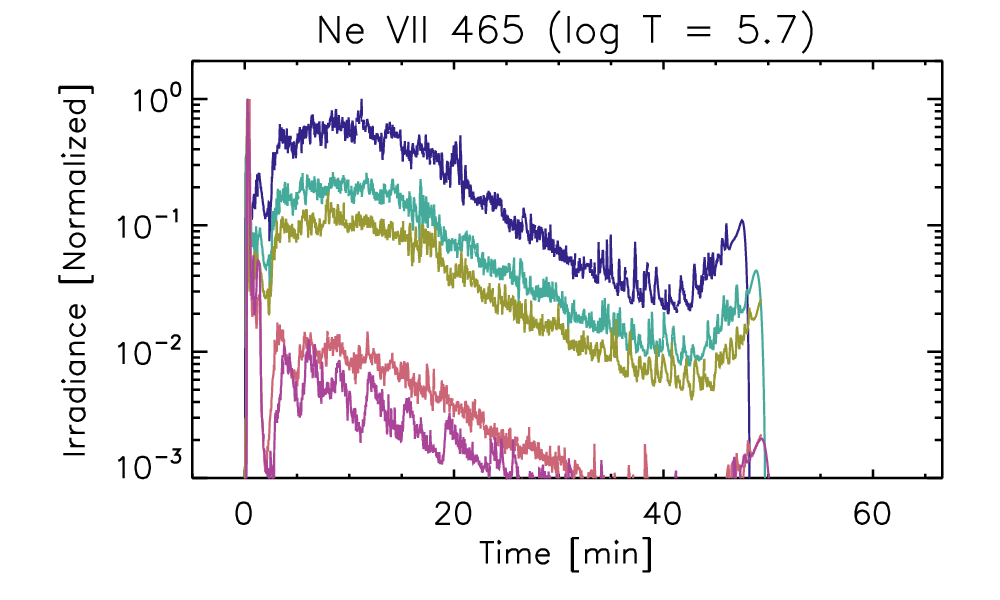}
    \caption{Irradiance time series for uniform, laminar 100 Mm loops, heated by an electron beam of $10^{11}$ erg s$^{-1}$ cm$^{-2}$ with various low-energy cutoffs $E_{c} = [10, 25, 30, 40, 50]$ keV.  The durations and intensities are strongly affected by the value of $E_{c}$, which affects the strength of evaporation.}
    \label{fig:Ec}
\end{figure*}

We next examine the low-energy cutoff $E_{c}$, which is the minimum value of the accelerated electron distribution (assuming a power law distribution of electrons, see discussions by \citealt{holman2011} and \citealt{xia2021}).  This value determines the average depth at which the electrons collide with the ambient plasma, and thus impacts the strength of chromospheric evaporation \citep{nagai1984}.  For smaller values of $E_{c}$, smaller energy fluxes can cause explosive evaporation, and vice versa \citep{reep2015}.  In Figure \ref{fig:Ec}, we show a comparison of five uniform, laminar loops heated by electron beams carrying the same total energy flux (peak of $10^{11}$ erg s$^{-1}$ cm$^{-2}$), with cutoffs $E_{c}$ of 10, 25, 30, 40, and 50 keV.  When the cutoff is relatively small, evaporation proceeds explosively, and thus the hottest lines are not greatly impacted by the value of $E_{c}$.  For large values of $E_{c}$, however, the evaporation is significantly weaker, and thus the corona is much lower density so that the intensities fade away much more quickly.  Additionally, for larger values of $E_{c}$, the transition region is not pushed significantly downwards in height, and thus the densities of the TR lines are not as high, reducing their intensities generally.  Finally, for large values of $E_{c}$, the energy deposition occurs too low to cause strong evaporation, but does still produce a spike in intensity during the heating period, which becomes relatively more prominent in cooler lines.  Compare the observational measurements in Appendix \ref{app:beams}, which shows the duration of SXR emission is weakly correlated with the low-energy cutoff.

\begin{figure*}
    \centering
    \includegraphics[width=0.48\textwidth]{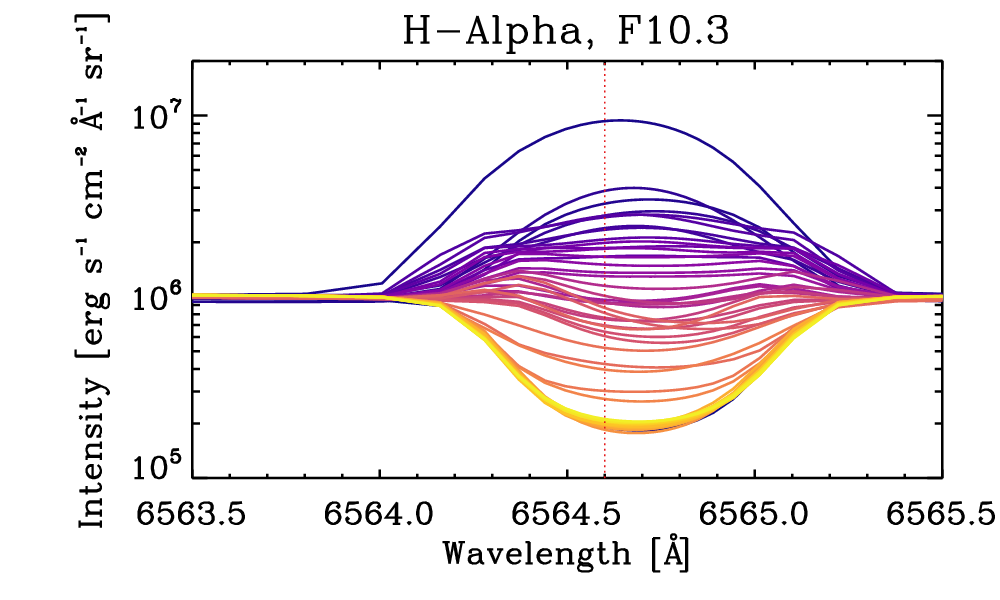}
    \includegraphics[width=0.48\textwidth]{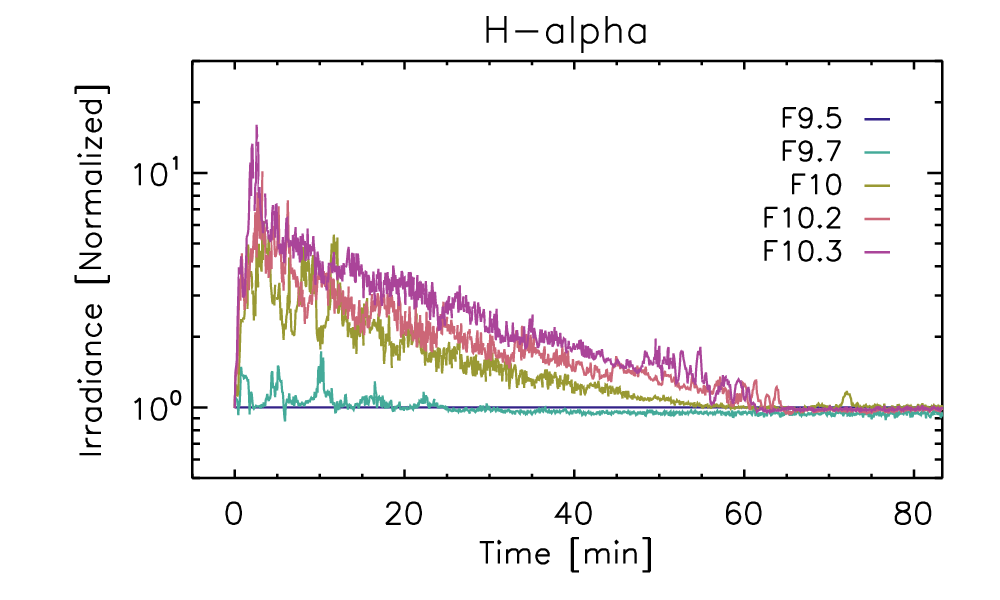}
    \includegraphics[width=0.48\textwidth]{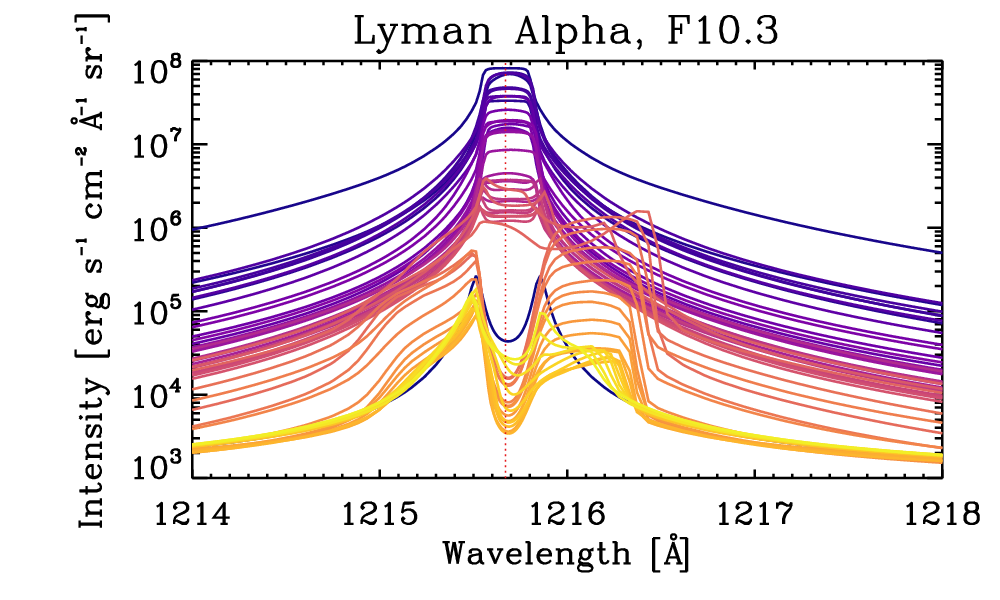}
    \includegraphics[width=0.48\textwidth]{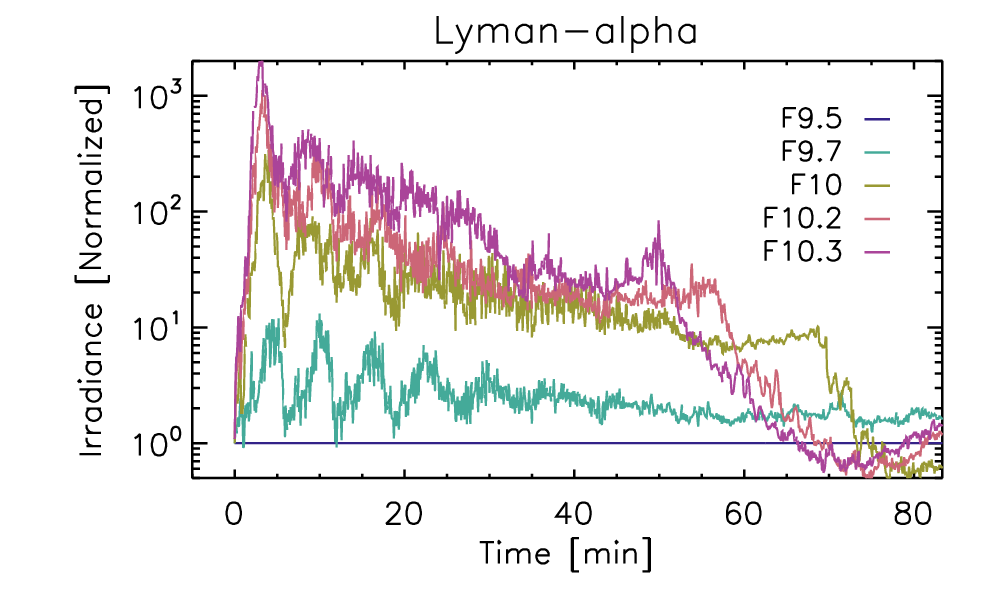}
    \includegraphics[width=0.48\textwidth]{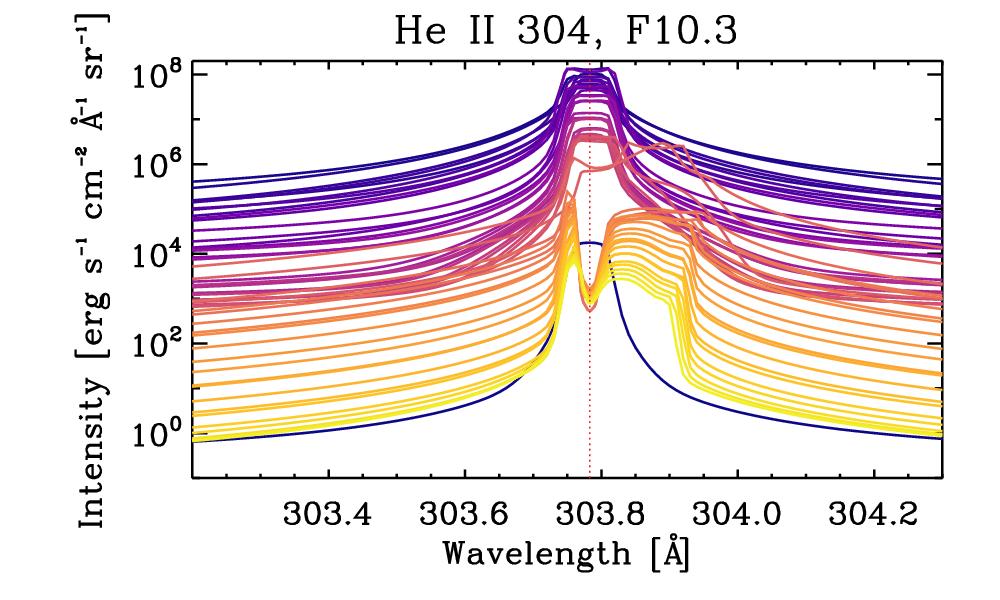}
    \includegraphics[width=0.48\textwidth]{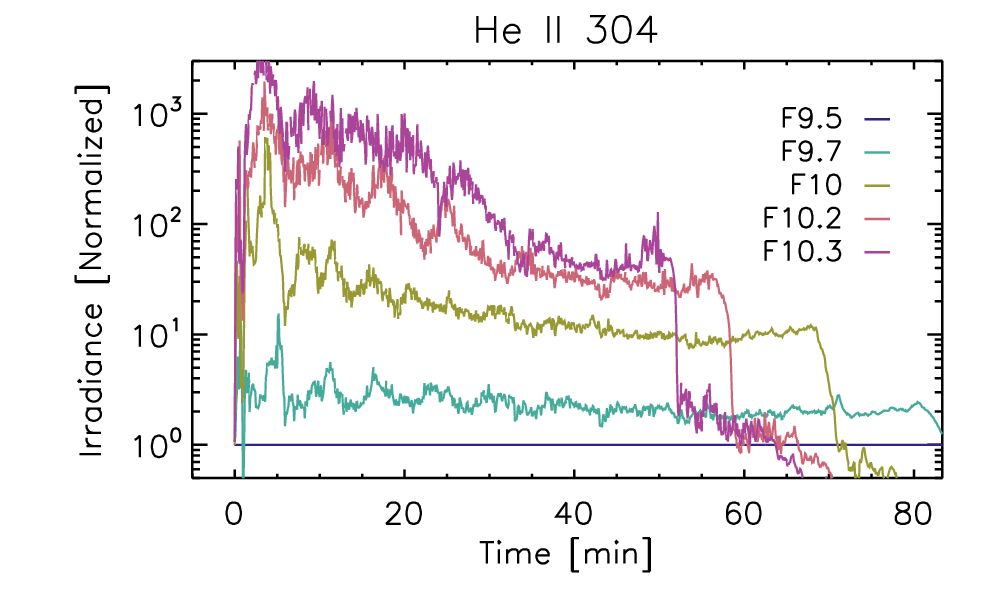}
    \caption{The evolution of H-$\alpha$, Lyman-$\alpha$, and \ion{He}{2} 304 \AA.  The plots at left show the evolution of the line profiles from a simulation with a peak heating rate of $F = 10^{10.3}$ erg s$^{-1}$ cm$^{-2}$, going from blue to yellow in time at a 100 s cadence, while the plots at right show the time series from five simulations with various heating rates.  }
    \label{fig:thick}
\end{figure*}

We finally examine the behavior a few lines that form deeper in the atmosphere, in the lower transition region and chromosphere.  In Figure \ref{fig:thick}, we show the evolution of H-$\alpha$, Lyman-$\alpha$, and \ion{He}{2} 304 \AA\ (Lyman-$\alpha$ for \ion{He}{2}).  The simulations were run on 100 Mm loops with low energy cut-off $E_{c} = 15$ keV and spectral index $\delta = 5$, with peak heating rates of $F = [10^{9.5}, 10^{9.7}, 10^{10.0}, 10^{10.2}, 10^{10.3}]$ erg s$^{-1}$ cm$^{-2}$.  The plots at left show the evolution of the line profiles for the strongest heating case, at a 100 s cadence from 0 to 5000 s from blue to yellow in time, where the dotted vertical line marks the rest wavelength (in vacuum).  The plots at right show the time series of the total irradiance for all five cases (normalized to the time zero value) at a 5 second cadence.  H-$\alpha$, forming in the mid-chromosphere \citep{carlsson2002}, is initially in absorption, quickly jumps into emission an order of magnitude above the continuum, and then gradually decays over time.  Stronger heating produces a longer duration time series in H-$\alpha$.  The behavior of Lyman-$\alpha$ and \ion{He}{2} 304, both forming in the upper chromosphere and lower transition region, behave somewhat differently.  In both cases, the initial central inversion disappears at the onset of heating, causing a strong brightening that decays over similar timescales as H-$\alpha$.  In both lines, there is a strong redshifted component during the gradual phase stemming from draining of material through the transition region.  In all three lines, the time series shows similar behavior to the optically thin transition region lines examine in previous Figures, with a long and gradual decay after a spike during the heating phase.  

\section{Discussion}

The evolution of a time series at a given wavelength depends intimately on the details of the heating, cooling, emission mechanism, and loop and arcade geometry.  Whether there is a correlation between the intensity and duration at a given wavelength depends primarily on the height in the atmosphere where the emission forms.  Cooler emission responds immediately to heating as the transition region is effectively pushed to lower heights in the atmosphere, which causes an immediate increase in the density and thus intensity.  On the other hand, hotter emission forming in the corona only sees a rise in density through chromospheric evaporation, which is delayed from the heating as it takes time to accelerate the plasma.  The strength of evaporation (speed or mass flux) depends on the exact heating parameters \citep{fisher1985,reep2015}, and therefore the rise in intensity is only an indirect response to the heating.  This is effectively the same as the Neupert effect \citep{neupert1968}, which is more traditionally stated as a relation between the HXR time series and time derivative of the SXRs.  More generally, it relates the heating to the evaporative response of the chromosphere, which causes that observed relation between HXRs and SXRs (\textit{e.g.} \citealt{li1993,lee1995,veronig2002}).  The Neupert effect is detected routinely in many observations in many different bandpasses of solar (\textit{e.g.} \citealt{dennis1993,kundu2001,lee2017}) and stellar flares (\citealt{osten2004,fuhrmeister2011,caballero2015}).

The Neupert effect points to a clear relation between the heating and observed emission.  The heating of individual loops within an arcade cannot be directly measured in general, and so must be inferred from other properties.  \citet{qiu2013} pointed out that there is a two-phase evolution in AIA's 1600 \AA\ bandpass (dominated by \ion{C}{4} emission) in a solar flare: a first phase that shows a direct response to heating, and a second phase that tracks the evaporation of material into the corona.  Observations of a two-phase evolution of time series have similarly been found in stellar flares (\textit{e.g.} \citealt{hawley2003}) and solar flares (\textit{e.g.} \citealt{zeng2014}).  These kinds of observations have also been used to attempt to constrain the heating properties of individual loops in a flaring arcade \citep{liu2013,zhu2018,qiu2021}, with the underlying assumption that brightenings were directly related to the heating.  \citet{rubio2016} similarly used individual peaks of the time derivative of GOES/XRS emission to inform heating functions in a flare arcade model.  

We summarize a few key points:
\begin{itemize}
    \item The Neupert effect implies different behavior of time series for hot and cold emission.  Heating causes the transition region to be pushed to lower heights, raising the density, which directly causes emission from cool plasma to brighten.  This is evidenced by the direct correlation between non-thermal HXR emission and transition region lines \citep{cheng1981,warren2016}.  Hot plasma brightens slowly as chromospheric evaporation raises the density of the coronal segment of loops in the flare arcade, which occurs more gradually.  
    \item The duration of emission in a given spectral line depends on the height in the atmosphere where it forms.  For hotter lines that form in the corona, the intensity in a loop rises with evaporation, peaks when the ionization fraction is maximized, and rapidly falls off below its formation temperature.  For cooler lines forming in the transition region or chromosphere, the emission peaks during the heating period when the density jumps up, stays roughly constant or slowly declines in the radiative cooling phase, and has a smaller secondary peak when the coronal portion of the loop cools to its formation temperature.  
    \item Cooling processes are vitally important for the duration of emission.  It is well known that longer loops take longer to both cool ($\propto L^{5/6}$) and drain ($\propto L$).  Expansion of the cross-sectional area suppresses draining of the loop, while turbulence suppresses cooling by thermal conduction, so these processes effectively lengthen the duration of light curves.  Both the area and the turbulence are vitally important, therefore, to correctly model loop evolution and time series, but are currently poorly constrained observationally.
    \item All of the parameters of the heating determine the temperatures and strength of evaporation.  Cooling due to radiation and conduction both depend intimately on the temperature and density, and thus the heating parameters indirectly impact the cooling durations.  Appendix \ref{app:beams} shows the relation of electron beam parameters to SXR duration, for example.
\end{itemize}

It is clear that the light curves of hot and cold emission evolve differently.  There is a test of the cartoon description in Figure \ref{fig:cartoon} in so-called ``cold'' solar flares, where there is ample evidence for particle acceleration in HXR or microwave emission emanating from the chromosphere, but the coronal temperature and emission measure are not raised significantly (\textit{e.g.} \citealt{fleishman2011,fleishman2016}).  A statistical study of 27 cold flares by \citet{lysenko2018} found that these events are typically shorter duration than the reference group in both HXRs and microwaves.  We have briefly examined the SXR emission for these 27 events in Appendix \ref{app:beams}.  We find that in cold flares, as with ``hot flares,'' there is no relation between SXR flux and duration, but the median duration ($\approx$ 90 s) is almost an order of magnitude shorter than the median duration of hot events ($\approx$ 630 s).  This suggests that there is only a very limited radiative cooling phase in cold flares, and the duration of the event corresponds roughly to the heating, but a more detailed study is warranted.  

In this work, we have only considered heating during the impulsive phase since we were considering single loops.  However, it is well known that there is significant energy release well into the gradual phase of flares \citep{czaykowska2001,isobe2002,qiu2013,kuhar2017}, which would, of course, change the duration of emission.  Additionally, we have only considered heating due to an electron beam, known to deliver significant amounts of heating to the chromosphere \citep{emslie2012}, but by no means the only mechanism of energy transport.  Other potential mechanisms are less well constrained, but likely include direct energy transport via thermal conduction (\textit{e.g.} \citealt{lopez2022}), wave generation of MHD waves (\textit{e.g.} \citealt{emslie1982,reep2016}), and acceleration of ion beams (\textit{e.g.} \citealt{canfield1985,kerr2023b}).  Future work with a multi-threaded model of flares should attempt to better constrain these mechanisms and their direct role in energy transport over the course of an event.

In summary, the duration of solar and stellar flares depends intimately on the details of the energy release and transport mechanisms.  The parameters of both the heating and cooling processes impact the duration of a flare.  In this paper, we have examined only flare heating mediated by an electron beam as HXR emission constraints its parameters, but there is certainly energy transport by ion beams and magnetohydrodynamic waves, which remains poorly constrained.  This is particularly important for the energy transport by larger electron energy fluxes greater than 10$^{12}$ erg cm$^{-2}$ s$^{-1}$, when electron beams become unstable due to excitation of Langmuir and ion-acoustic waves \citep{drummond1970,zharkova2011,yao2022}.  Powerful stellar superflares at energies over 10$^{34}$ erg occur in larger active regions, with stronger magnetic flux, and thus more efficiently accelerate particles, including both electron and ion beams.  As shown by \citet{zharkova2011}, intense electron beams are expected to be damped in higher layers of the chromosphere, causing excitation of plasma waves.  Meanwhile, intense proton beams are stable with respect to excitation of collective modes in plasma and their mean free path is mostly driven by the collisional dissipation of their energy, and thus can reach deeper, denser layers of the stellar chromosphere, causing significant line broadening due to the Stark effect \citep{kowalski2017}.  All of these factors play a role on the energy transport through the lower atmosphere and the resultant impacts on X-ray spectra \citep{alaoui2017,allred2020,alaoui2021}.  In order to better understand the impact of stellar flares on (exo)planets, more work is needed to build a full thermodynamic model of flare irradiance across the X-ray, extreme, far, and near ultraviolet bands to simulate the response of their magnetosphere, upper and lower atmospheres.

\nopagebreak

\leavevmode \newline

\acknowledgments  
JWR was supported by the Office of Naval Research 6.1 Support Program.  VSA was supported by the GSFC Sellers Exoplanet Environments Collaboration (SEEC), which is funded by the NASA Planetary Science Division’s Internal Scientist Funding Model (ISFM) and NASA NNH21ZDA001N-XRP F.3 Exoplanets Research Program grants.  
We thank Dr. Konstantin Getman for helpful discussions.  The authors also thank the two anonymous referees for suggestions which have improved this manuscript.

\appendix

\section{Statistical Relations}
\label{app:beams}
\begin{figure*}
\centering
\includegraphics[width=0.42\textwidth]{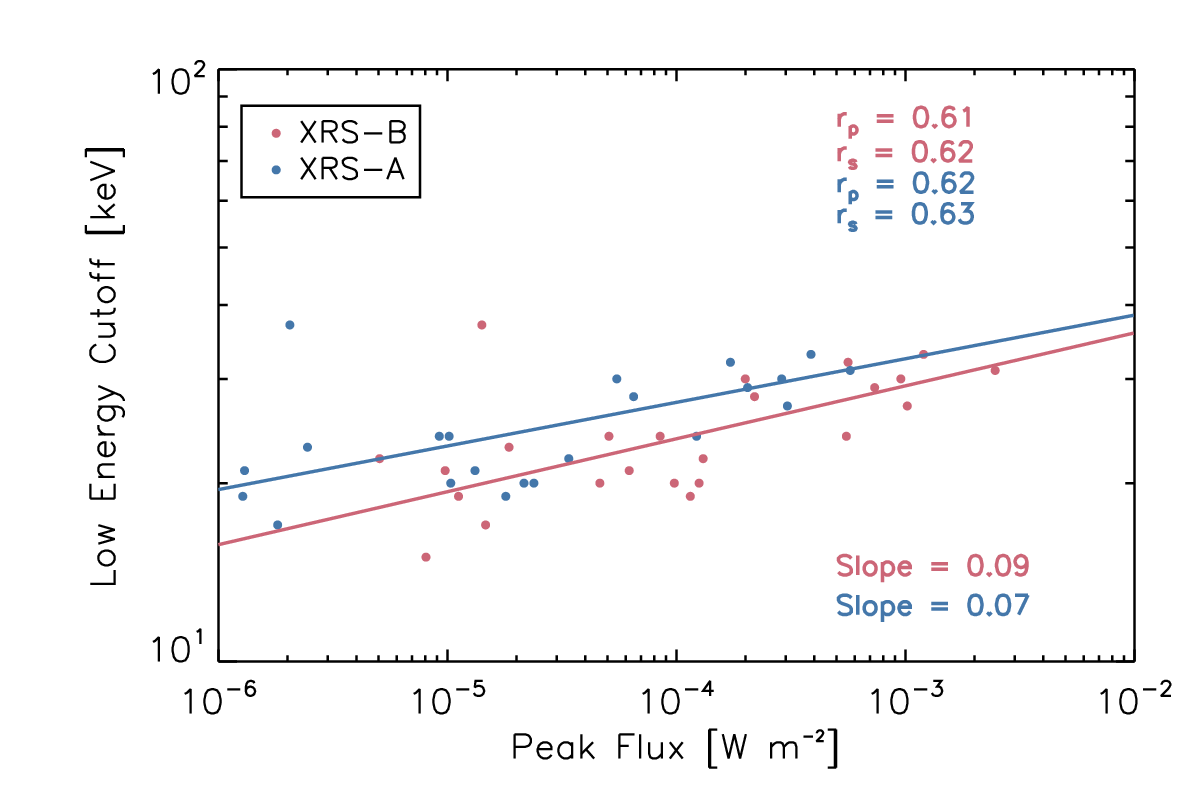}
\includegraphics[width=0.42\textwidth]{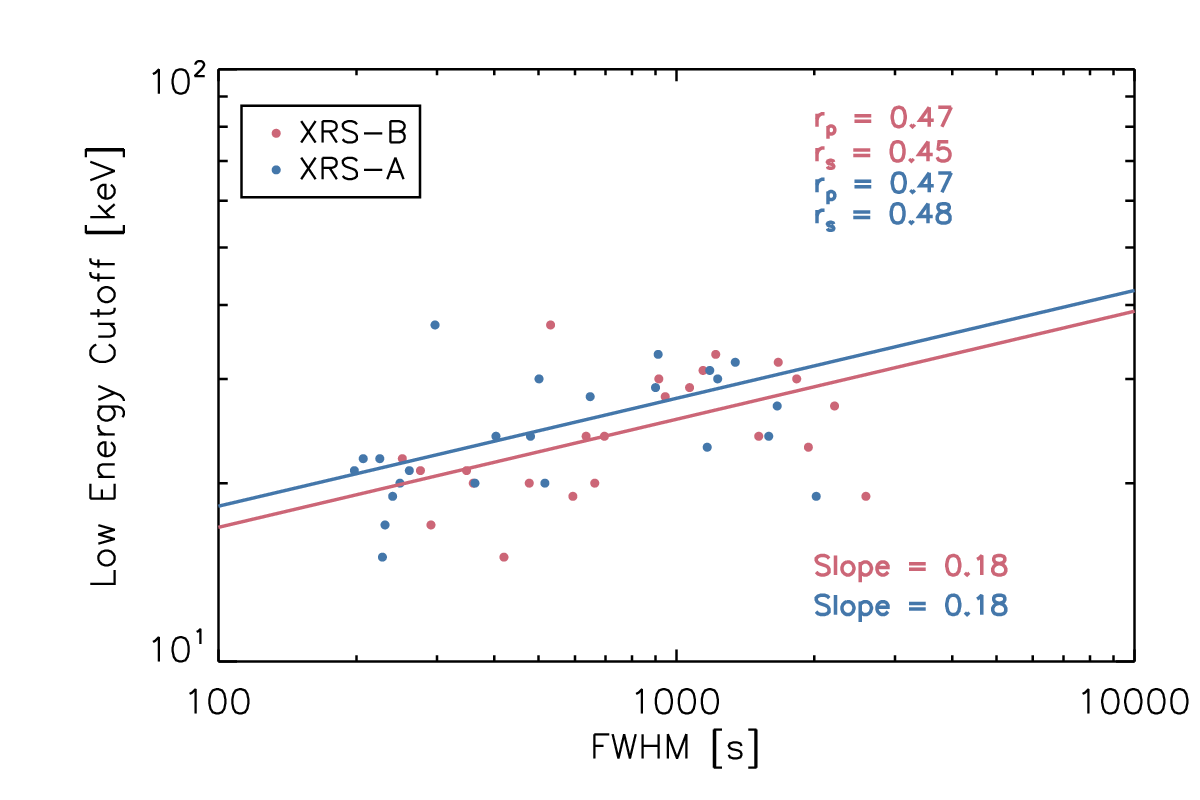}
\includegraphics[width=0.42\textwidth]{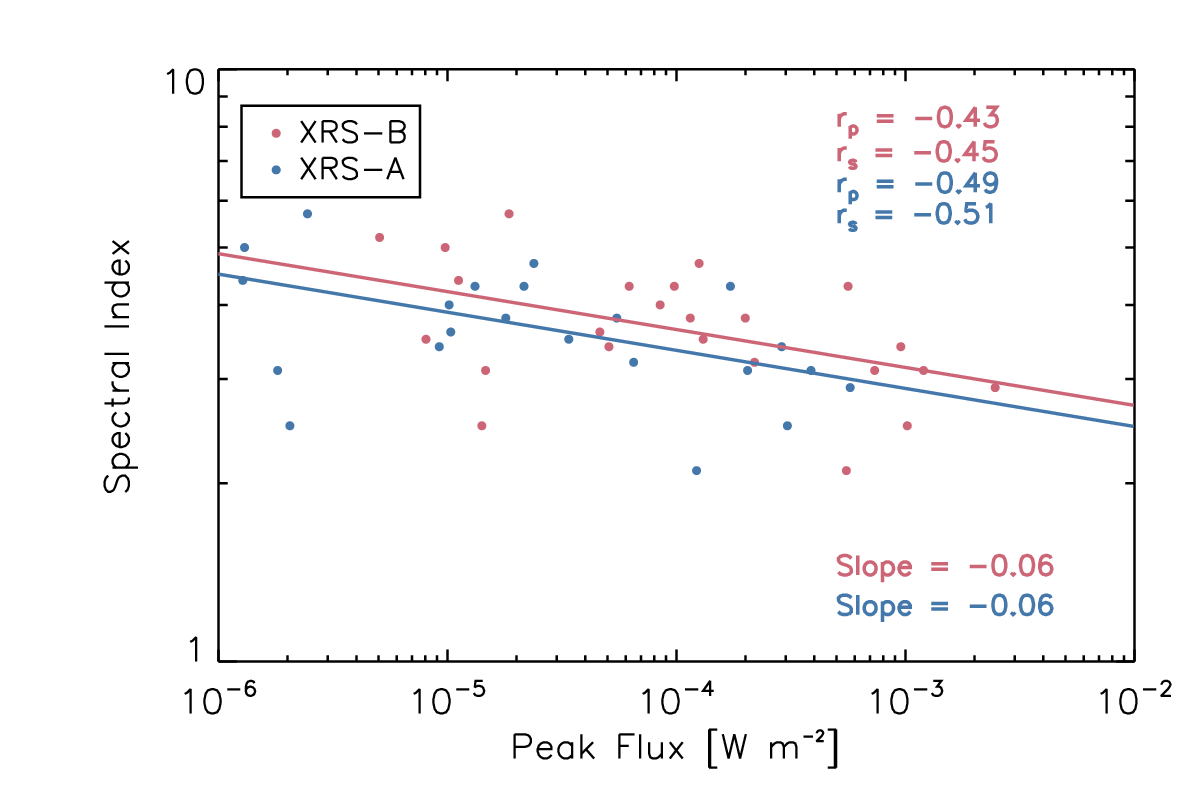}
\includegraphics[width=0.42\textwidth]{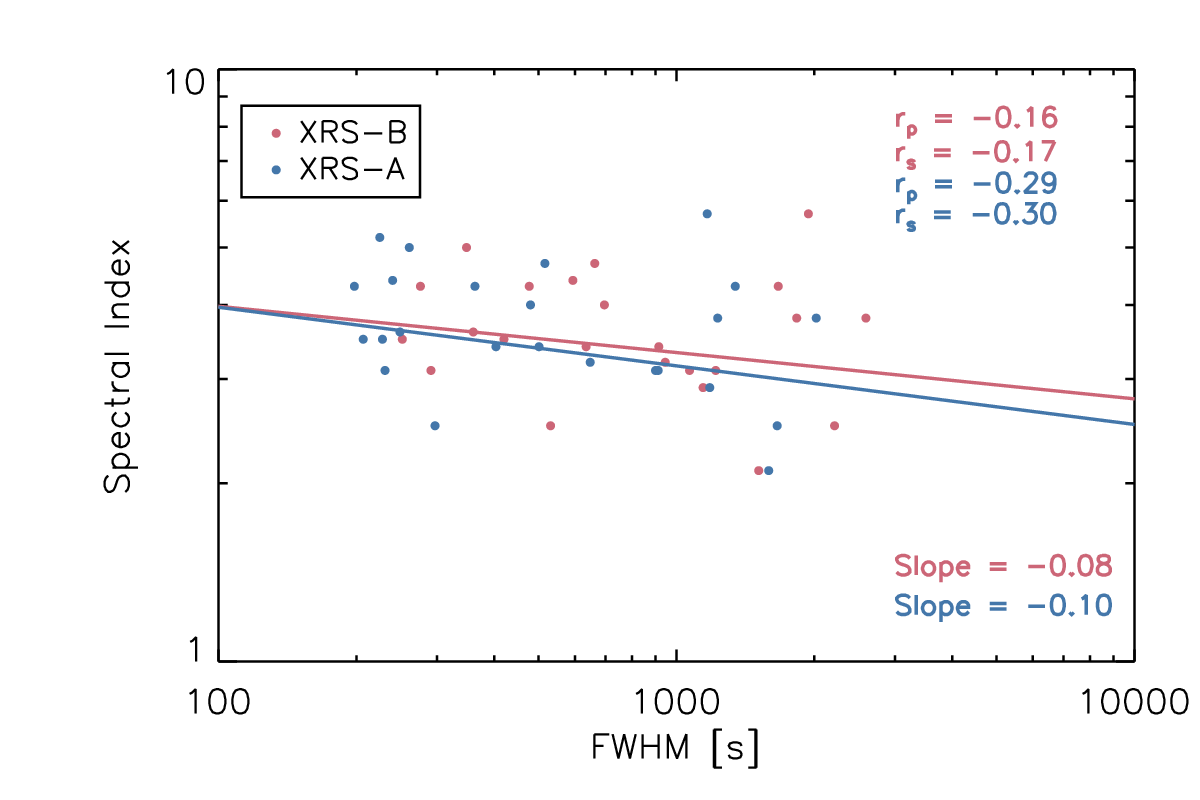}
\includegraphics[width=0.42\textwidth]{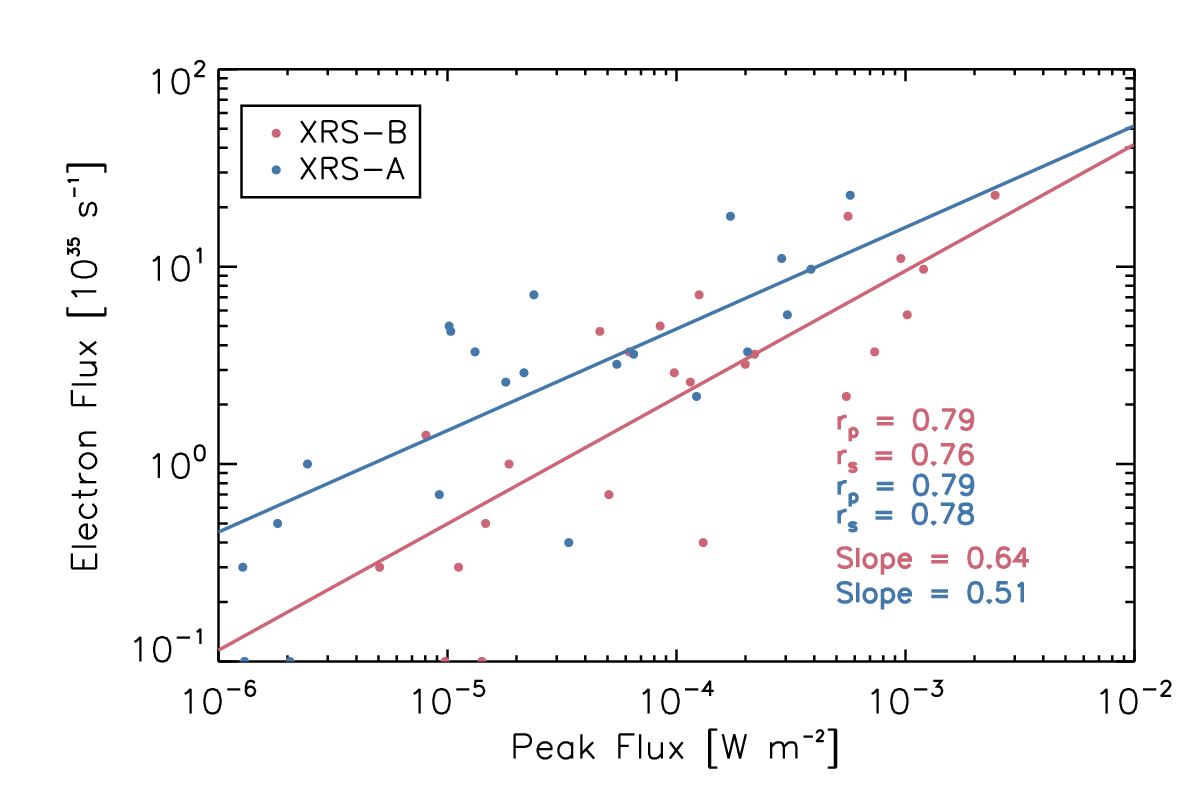}
\includegraphics[width=0.42\textwidth]{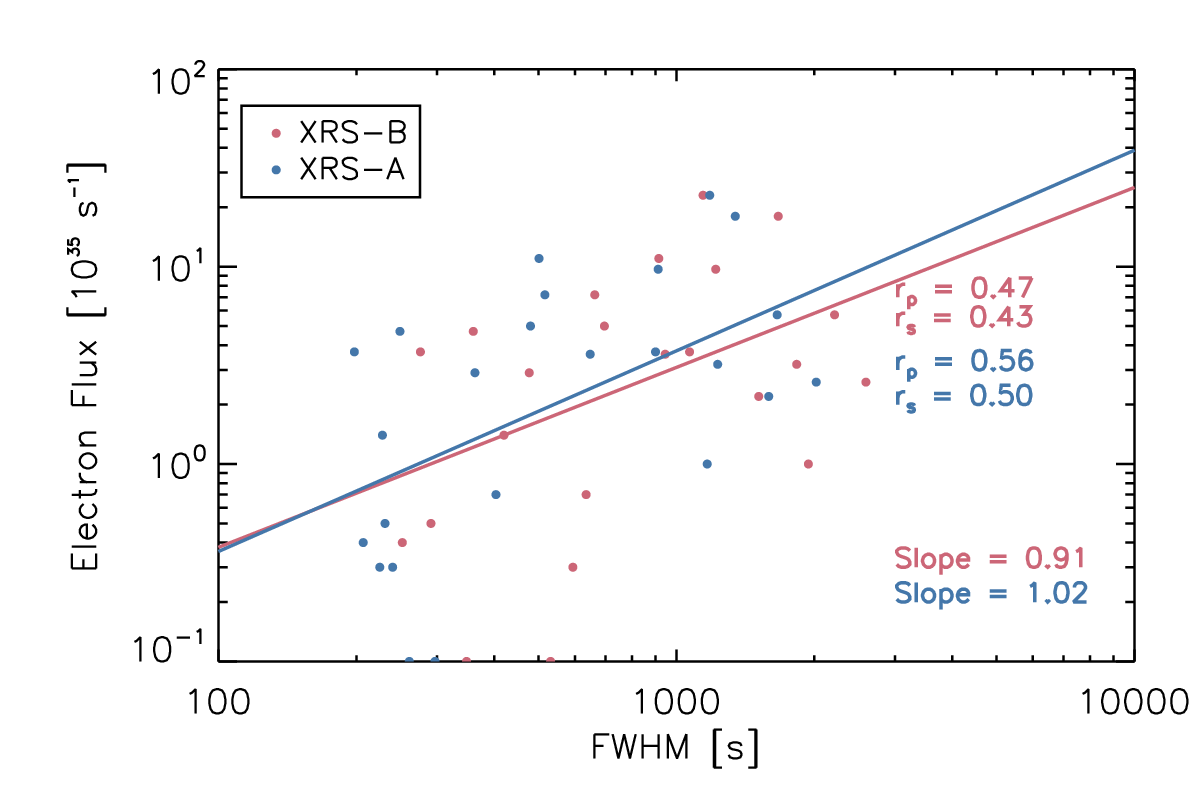}
\includegraphics[width=0.42\textwidth]{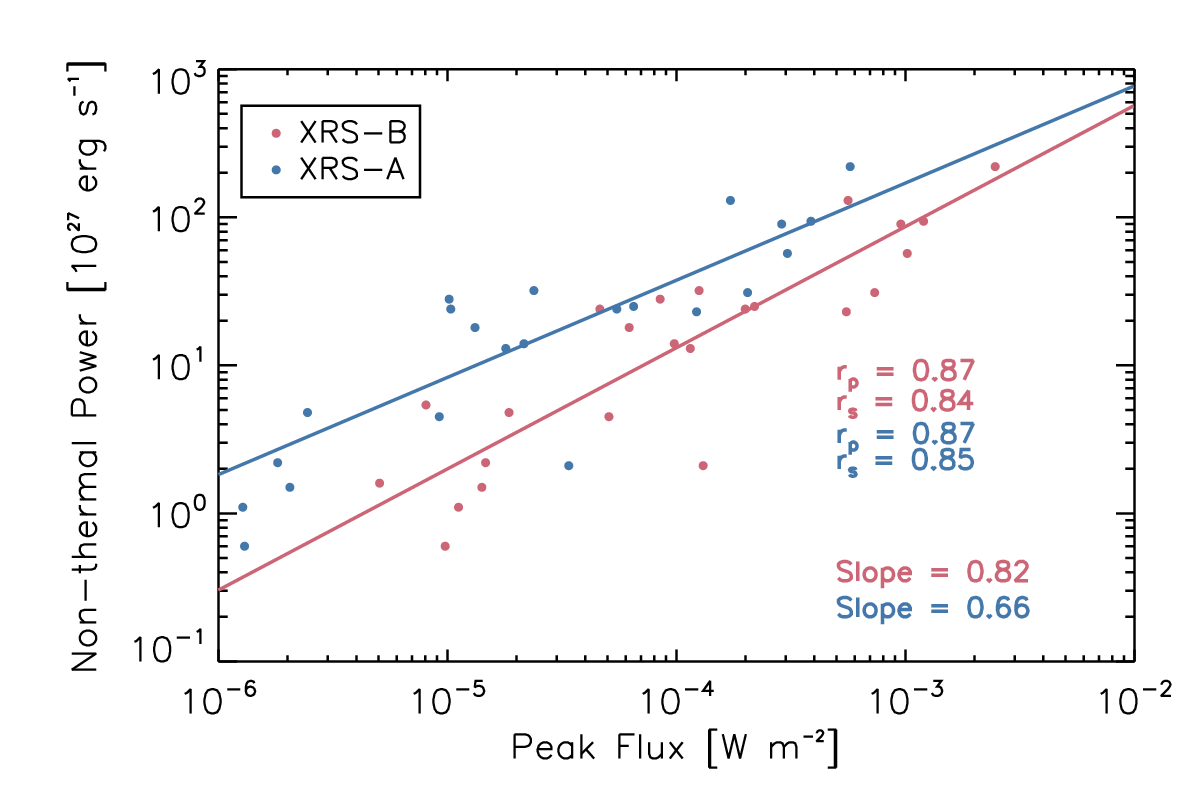}
\includegraphics[width=0.42\textwidth]{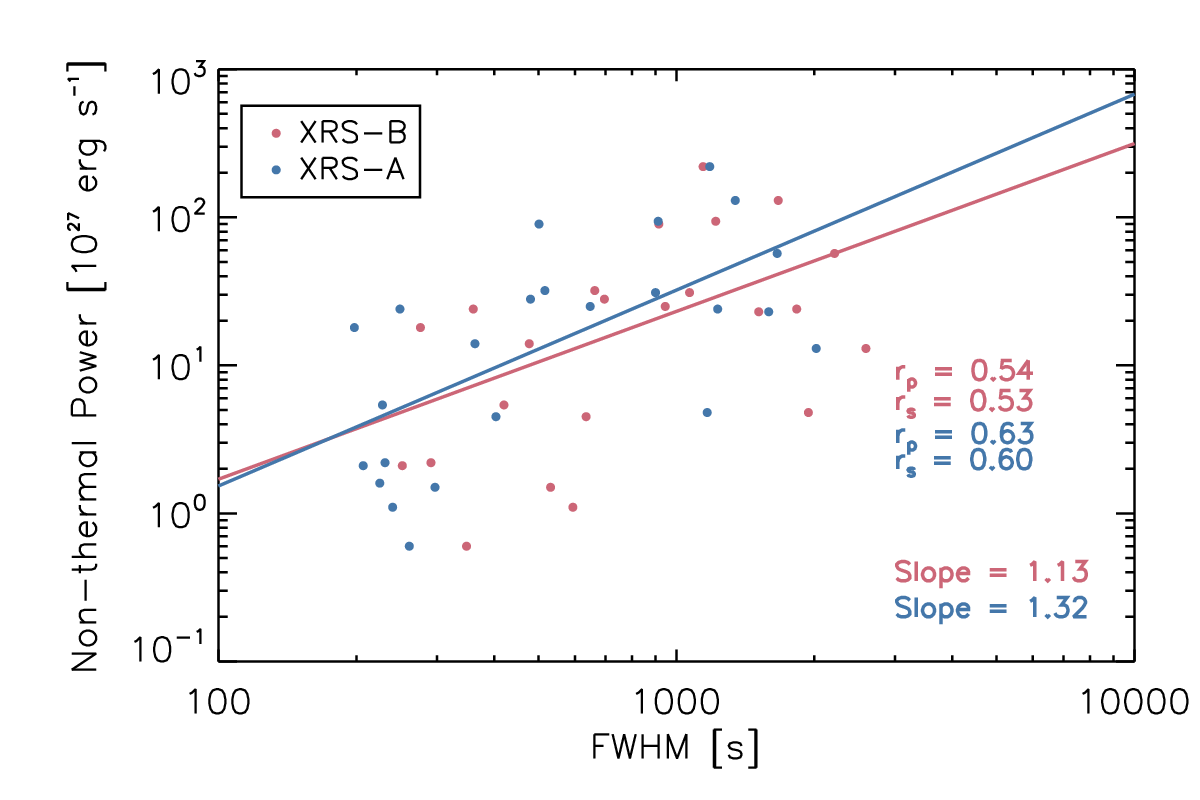}
\caption{Scatter plots comparing the electron beam parameters in 24 flares measured by \citet{warmuth2016a} to the XRS peak flux and the XRS FWHM (in each channel, XRS-A in blue, XRS-B in red).  The Pearson $r_p$ (correlation) and Spearman $r_s$ (monotonicity) coefficients are indicated.  As reported by those authors, the parameters are strongly correlated with the peak SXR flux.  Additionally, there may be a weaker correlation between the low-energy cut-off, electron rate, and non-thermal power with the duration of the flare in SXRs.  Theil-Sen linear fits in log-log space have been overplotted and their slopes labeled.  \label{fig:warmuth}}
\end{figure*}

We can check whether the SXR duration is related to the electron beam parameters.  \citet{warmuth2016a} measured these parameters using spectral inversion with RHESSI for a sample of 24 flares ranging from C to X-class.  We plot their measured values against the peak flux in both GOES/XRS channels as well as the FWHM of both channels for each of these events.  The scatter plots in Figure \ref{fig:warmuth} show this comparison.  The plots also note the Pearson correlation coefficient $r_{p}$ and Spearman coefficient $r_{s}$ for both XRS channels.  The lines show a linear regression in log-log space done with a non-parametric Theil-Sen estimator, which is resistant to outliers.  Their measured slopes are also indicated.  As noted by \citet{warmuth2016a}, there is a clear correlation between the XRS-B flux and each of the electron beam parameters (low-energy cut-off, spectral index, electron rate, and non-thermal power).  This holds in both XRS channels.  Additionally, there is a weaker correlation between the duration in each XRS channel with the low-energy cut-off, electron rate, and non-thermal power, but perhaps not with the spectral index.

\begin{figure*}
\centering
\includegraphics[width=0.32\textwidth]{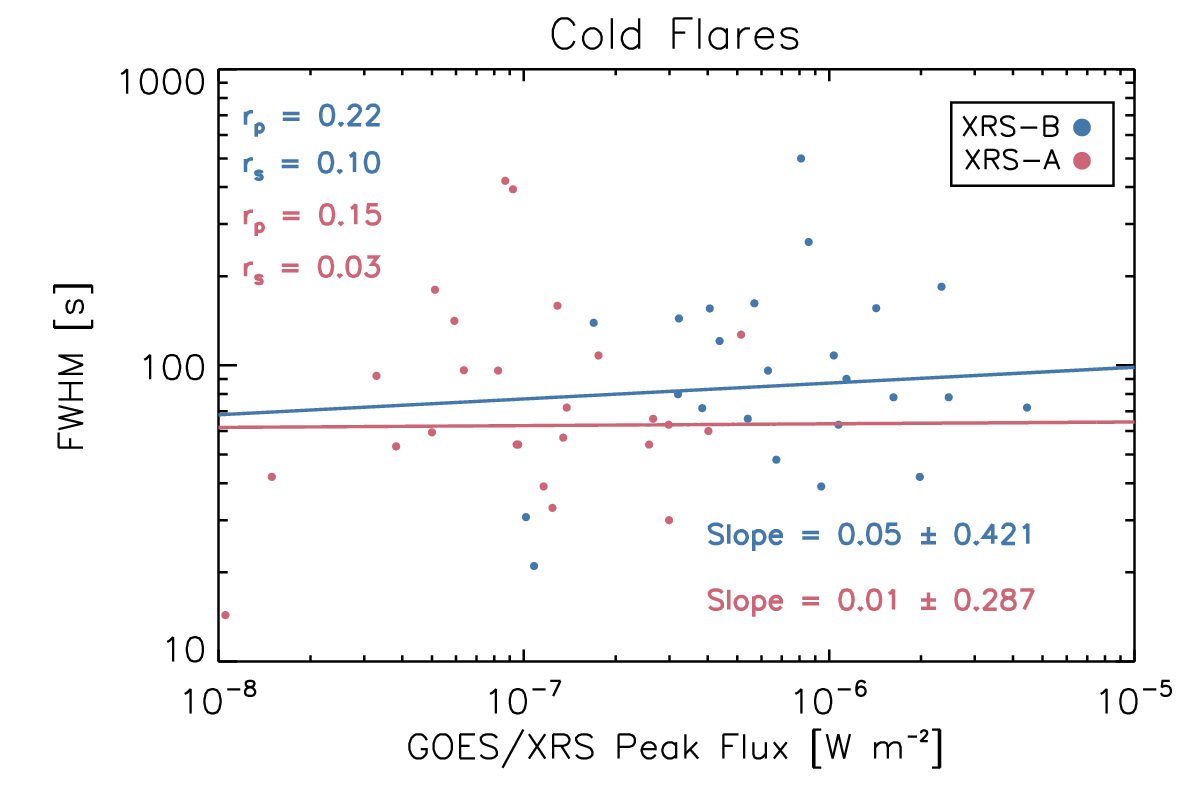}
\includegraphics[width=0.32\textwidth]{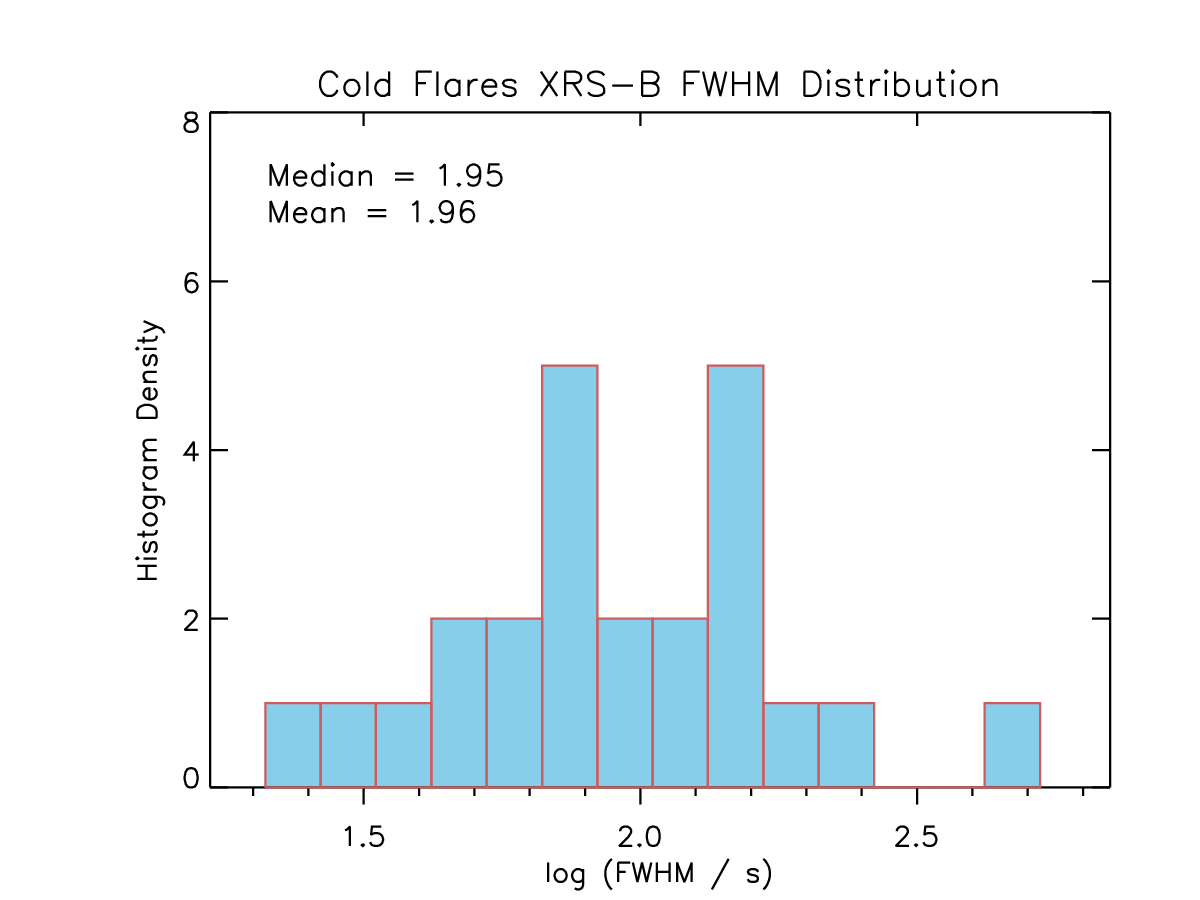}
\includegraphics[width=0.32\textwidth]{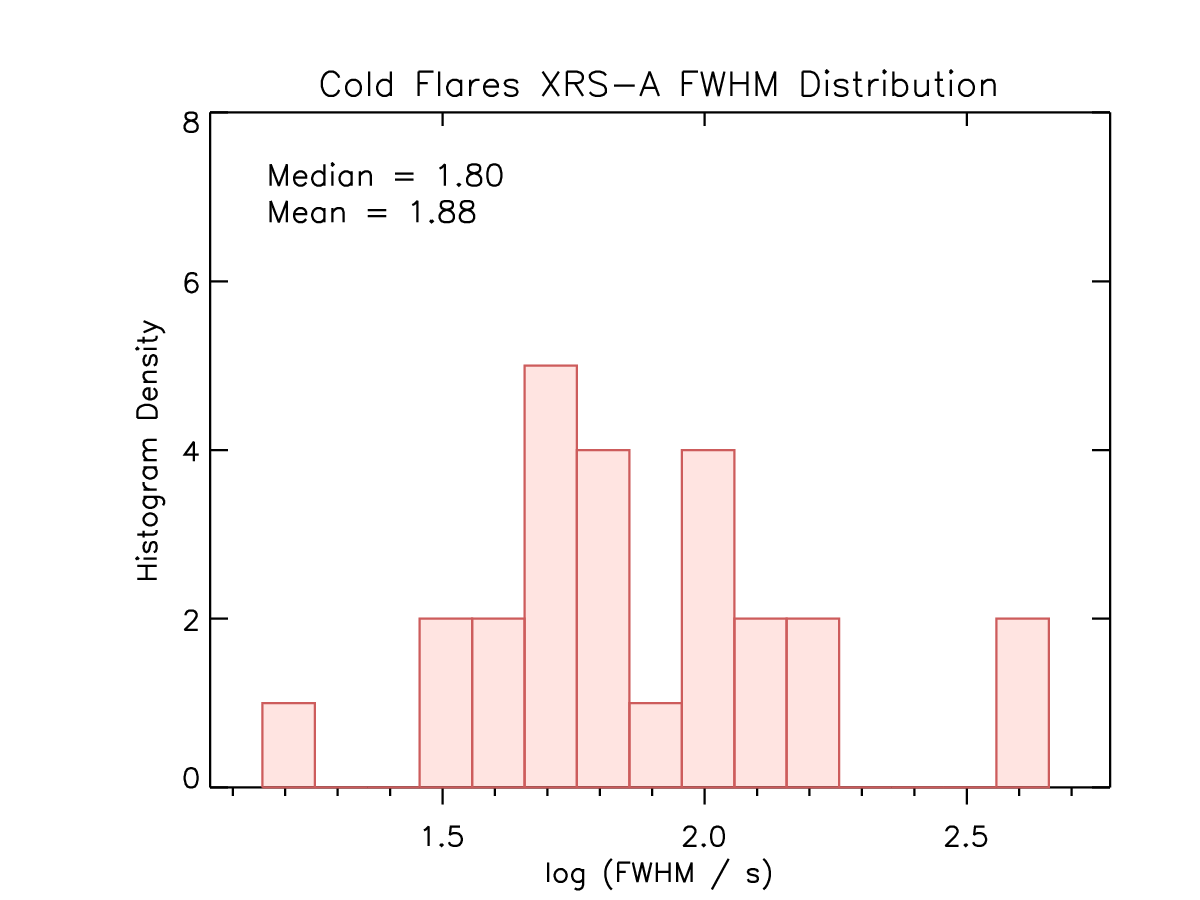}
\caption{A scatter plot of peak flux and duration in each XRS channel for the set of 27 cold flares from \citet{lysenko2018}.  The correlation coefficients are small, and the slopes are consistent with 0. The distributions of durations are also shown for each XRS channel.  The average durations are about an order of magnitude shorter than for ``hot'' flares.  \label{fig:cold}} 
\end{figure*}

We have also examined the durations of so-called cold flares, where there is noticeable emission from non-thermal bremsstrahlung (in either HXRs or radio), but little increase in temperature or emission measure.  In these events, we expect that the SXR emission should be impulsive, responding to the heating and not significantly to the more gradual evaporation.  \citet{lysenko2018} presented a statistical study of 27 such events in depth.  Using the event list in that paper, in Figure \ref{fig:cold} we have plotted the background-subtracted SXR flux against SXR duration, finding little correlation (small Pearson $r_{p}$ and small Spearman $r_{s}$).  The Theil-Sen linear fits have a slope consistent with 0 in both channels.  Additionally, we show the distributions of FWHM in each channel, with median values of 63 and 90 s in XRS-A and XRS-B, respectively.  The statistics are too limited to make any conclusions about the distribution.  \citet{reep2019} showed similar distributions for around 3,000 ``hot'' flares observed with GOES-15, finding median values about 360 and 630 s in XRS-A and XRS-B.  In both hot and cold flares, there is no relation between duration and GOES class, but cold flares are of notably shorter duration, indicating that the radiative cooling phase is negligible in cold flares.

\bibliography{apj}
\bibliographystyle{aasjournal}

\end{document}